\documentclass[aos,MSNbibl,seceqn,nameyear,rotating,dvips]{arximspdf}
\usepackage{dcolumn}

\doi{10.1214/12-AOS1024} 
\volume{40}
\issue{3}
\pubyear{2012}
\firstpage{1846}
\lastpage{1877}

\makeatletter
\newcommand{\sset}[1]{,\ (#1)}

\newcolumntype{d}[1]{D{.}{.}{#1}}
\newcolumntype{k}[1]{D{,}{}{#1}}

\newproclaim{Exam}{Example}

\newproclaim{rema}{Remark}
\newtheorem{theo}{Theorem}
\newtheorem{prop}{Proposition}

\newcommand{\E}{\mathrm{E}}

\newcommand{\bbeta}{\bolds\beta}

\newcommand{\bfX}{\mathbf{X}}
\newcommand{\bfY}{\mathbf{Y}}

\newcommand{\bX}{\mathbf{X}}
\newcommand{\bomega}{\bolds{\omega}}
\newcommand{\bvarepsilon}{\bolds{\varepsilon}}

\makeatother

\begin{document}
\begin{frontmatter}

\title{Robust rank correlation based screening}
\runtitle{Robust rank correlation based screening}

\begin{aug}
\author[A]{\fnms{Gaorong} \snm{Li}\thanksref{m1}\ead[label=e1]{ligaorong@gmail.com}},
\author[B]{\fnms{Heng} \snm{Peng}\corref{}\thanksref{m2}\ead[label=e2]{hpeng@math.hkbu.edu.hk}},
\author[C]{\fnms{Jun} \snm{Zhang}\thanksref{m3}\ead[label=e3]{zhangjunstat@gmail.com}}
\and
\author[B]{\fnms{Lixing} \snm{Zhu}\thanksref{m4}\ead[label=e4]{lzhu@math.hkbu.edu.hk}}
\runauthor{Li, Peng, Zhang and Zhu}
\affiliation{Beijing University of Technology, Hong
Kong Baptist University, Shenzhen~University and
Hong Kong Baptist University}
\address[A]{G. Li\\
College of Applied Sciences\\
Beijing University of Technology\\
Beijing 100124\\
China\\
\printead{e1}}
\address[B]{H. Peng\\
L. Zhu\\
Department of Mathematics\\
Hong Kong Baptist University\\
Hong Kong, China\\
\printead{e2}\\
\phantom{E-mail: }\printead*{e4}}
\address[C]{J. Zhang\\
Shen Zhen-Hong Kong Joint Research Center\\
\quad for Applied Statistics\\
Shenzhen University\\
Shenzhen 518060\\
China\\
\printead{e3}} 
\end{aug}

\thankstext{m1}{Supported in part by the NNSF (11101014) of China,
the Specialized Research Fund for the Doctoral Program of Higher
Education of China (20101103120016), PHR (IHLB, PHR20110822),
the Training Programme Foundation for the Beijing Municipal
Excellent Talents (2010D005015000002) and the
Fundamental Research Foundation of Beijing University of Technology
(X4006013201101).}

\thankstext{m2}{Supported in part by CERG grants from the Hong Kong Research Grants Council
(HKBU 201610, HKBU 201809 and HKBU 202012), FRG grants from Hong Kong Baptist University
(FRG/10-11/024 and FRG/11-12/130) and a grant from National Nature Science Foundation of China (NNSF 11271094).}

\thankstext{m3}{Supported in part by the NNSF
(11101157) of China.}

\thankstext{m4}{Supported in part by a grant from the
Research Grants Council of Hong Kong,
and an FRG grant from Hong Kong Baptist University.}

\received{\smonth{3} \syear{2012}}
\revised{\smonth{6} \syear{2012}}

%
\begin{abstract}
Independence screening is a variable selection method that uses a
ranking criterion to select significant variables, particularly for
statistical models with nonpolynomial dimensionality
or ``large $p$, small $n$'' paradigms when $p$
can be as large as an exponential of the sample size $n$. In this
paper we propose a robust rank correlation screening (RRCS) method
to deal with ultra-high dimensional data. The new procedure is based
on the Kendall $\tau$ correlation coefficient between response and
predictor variables rather than the Pearson correlation of existing
methods. The new method has four desirable features compared with
existing independence screening methods. First, the sure
independence screening property can hold only under the existence of
a second order moment of predictor variables, rather than
exponential tails or alikeness, even when the number of predictor
variables grows as fast as exponentially of the sample size. Second,
it can be used to deal with semiparametric models such as
transformation regression models and single-index models under
monotonic constraint to the link function without involving
nonparametric estimation even when there are nonparametric functions
in the models. Third, the procedure can be largely used against
outliers and influence points in the observations. Last, the use of
indicator functions in rank correlation screening greatly simplifies
the theoretical derivation due to the boundedness of the resulting
statistics, compared with previous studies on variable screening.
Simulations are carried out for comparisons with existing methods
and a real data example is analyzed.
\end{abstract}

%

\end{frontmatter}

\section{Introduction}\label{sec1}

With the development of scientific techniques, ultra-high dimensional
data sets have appeared in diverse areas of the sciences, engineering
and humanities; \citet{Don00} and \citet{FanLi06} have
provided comprehensive reviews. To handle statistical problems related
to high dimensional data, variable/model selection plays an important
role in establishing working models that include significant variables
and exclude as many insignificant variables as possible. A very
important and popular methodology is shrinkage estimation with
penalization, with examples given of bridge regression
[\citet{FraFri93}, \citet{HuaHorMa08}], LASSO
[\citet{Tib96}, \citet{van08}], elastic-net
[\citet{ZouHas05}], adaptive LASSO [\citet{Zou06}], SCAD
[\citet{FanLi01,FanPen04,FanLv11}] and
Dantzig selector [\citet{CanTao07}]. When irrepresentable
conditions are assumed, we can guarantee selection consistency for
LASSO and Dantzig selector even for ``large $p$, small $n$'' paradigms
with nonpolynomial dimensionality (NP-dimensionality). However,
directly applying LASSO or Dantzig selector to ultra-high dimensional
modeling is not a good choice because the irrepresentable conditions
can be rather stringent in high dimensions; see, for example,
\citet{LvFan09} and \citet{FanLv10}.

\citet{FanLv08} proposed another promising approach called sure
independence screening (SIS). This methodology has been developed in
the literature by researchers recently. \citet{FanSon10}
extended SIS to ultra-high dimensional generalized linear models,
and \citet{FanFenSon11} studied it for ultra-high dimensional
additive models. Moreover, based on the idea of dimension
reduction, \citet{Zhuetal11} suggested a model-free feature
screening method for most generalized parametric or semiparametric
models. To sufficiently use the correlation information among the
predictor variables, \citet{Wan12} proposed a factor profile sure
screening method for the ultra-high dimensional linear regression
model. Different from existing methods with penalization, SIS does
not use penalties to shrink estimation, but ranks the importance of
predictors by correlations between response and predictors
marginally for variable/model selection. To perform the ranking,
Pearson correlation is adopted; see \citet{FanLv08}. For
NP-dimensionality, the tails of predictors need to be
nonpolynomially light. This is also the case for other shrinkage
estimation methods such as the LASSO and Dantzig selector. Moreover,
to use more information among the predictor variables to make a sure
screening such as \citet{Wan12}, or to apply the sure screening
method to more general statistical models such as
\citet{Zhuetal11}, more restrictive conditions, such as the normality
assumption [\citet{Wan12}] or the linearity and moment conditions
[\citet{Zhuetal11}], need be imposed on the predictor variables. To
further improve estimation efficiency, \citet{FanLv08} suggested a
two-stage procedure. First, SIS is used as a fast but crude method
of reducing the ultra-high dimensionality to a relatively large
scale that is smaller than or equal to the sample size $n$; then, a
more sophisticated technique can be applied to perform the final
variable selection and parameter estimation simultaneously. Note
that for linear models, the SIS procedure also depends on the
explicit relationship between the Pearson correlation and the least
squares estimator [\citet{FanLv08}]. For generalized linear models,
\citet{FanSamWu09} and \citet{FanSon10} selected
significant predictors by sorting the corresponding marginal
likelihood estimator or marginal likelihood. That method can be
viewed as a likelihood ratio screening, as it builds on the
increments of the log-likelihood. The rate of $p$ also depends on
the tails of predictors. The lighter the tails are, the faster the
rate of $p$ can be. \citet{XuZhu} also showed for longitudinal
data that when only the moment condition is assumed, the rate of $p$
cannot exponentially diverge to infinity unless moments of all
orders exist.

For other semiparametric models such as transformation models and
single-index models, existing SIS procedures may involve
nonparametric plug-in estimation for the unknown transformation or
link function. This plug-in may deteriorate the estimation/selection
efficiency for NP-dimensionality problems. Although the innovative
sure screening method proposed by \citet{Zhuetal11} can be
applied to more general parametric or semiparametric models, as
commented above, the much more restrictive conditions are required
for the predictor variables. \citet{Zhuetal11} imposed some
requirements for the tail of the predictor variables which further
satisfy the so-called linearity condition. This condition is only
slightly weaker than elliptical symmetry of the distribution of the
predictor vector [\citet{Li91N2}]. It is obvious that their sure screening
method does not have the robust properties as the proposed method in
this paper has. Further, when the categorial variables do involve the
ultra-high dimensional predictor vector, the restrictive
conditions on the predictor variables hinder the model-free feature
screening method to apply directly. On the other hand, such a
model-free feature screening method is based on slice inverse
regression [SIR, \citet{Li91N2}]. It is well known that SIR is not workable
to the model with symmetric regression function; see \citet{CoWe91}.

We note that the idea of SIS is based on Pearson correlation learning.
However, the Pearson correlation is not robust against heavy tailed
distributions, outliers or influence points, and the nonlinear
relationship between response and predictors cannot be discovered by
the Pearson correlation. As suggested by \citet{HalMil09} and
\citet{HuaHorMa08}, independence screening could be conducted with
other criteria. For correlation relationships, there are several
measurements in the literature, and the Kendall $\tau$
[\citet{Ken38}] is a very commonly used one that is a correlation
coefficient in a nonparametric sense. Similar to the Pearson
correlation, the Kendall $\tau$ also has wide applications in
statistics. \citet{Ken62} gave an overview of its applications in
statistics and showed its advantages over the Pearson correlation.
First, it is robust against heavy tailed distributions: see
\citet{Sen68} for parameter estimation in the linear regression
model. Second, the Kendall $\tau$ is invariant under monotonic
transformation. This property allows us to discover the nonlinear
relationship between the response and predictors. For example,
\citet{Han87} suggested a maximum rank correlation estimator (MRC)
for the transformation regression model with an unknown transformation
link function. Third, the Kendall $\tau$ based estimation is a
U-statistic with a bounded kernel function, which provides us a chance
to obtain sure screening properties with only a moment condition.
Another rank correlation is the Spearman correlation [see, e.g.,
\citet{WacMenSch}]. The Spearman rank
correlation coefficient is equivalent to the traditional linear
correlation coefficient computed on ranks of items [\citet{WacMenSch}]. The Kendall $\tau$ distance between
two ranked lists is proportional to the number of pairwise adjacent
swaps needed to convert one ranking into the other. The Spearman rank
correlation coefficient is the projection of the Kendall $\tau$ rank
correlation to linear rank statistics. The Kendall $\tau$ has become a
standard statistic with which to compare the correlation between two
ranked lists. When various methods are proposed to rank items, the
Kendall $\tau$ is often used to measure which method is better relative
to a ``gold standard.'' The higher the correlation between the output
ranking of a method and the ``gold standard,'' the better the method
is. Thus, we focus on the Kendall $\tau$ only. More interestingly, the
Kendall $\tau$ also has a close relationship with the Pearson
correlation, particularly when the underlying distribution of two
variables is a bivariate normal distribution (we will give the details
in the next section). As such, we can expect that a Kendall $\tau$
based screening method will benefit from the above mentioned advantages
to be more robust than the SIS.

The reminder of this paper is organized as follows. In Section
\ref{sec2} we give the details of the robust rank correlation
screening method (RRCS) and present its extension to ultra-high
dimensional transformation regression models. In Section
\ref{sec3} the screening properties of the RRCS are studied
theoretically for linear regression models and transformation
regression models. In Section~\ref{sec4} an iterative RRCS
procedure is presented. We also discuss RRCSs application to
generalized linear models with NP-dimensionality. Numerical studies
are reported in Section~\ref{sec5} with a comparison with the SIS.
Section~\ref{sec7} concludes the paper. A real example and the
proofs of the main results can be found in the supplementary
material for the paper [\citet{Lietal}].

\section{Robust rank correlation screening (RRCS)}\label{sec2}
\subsection{\texorpdfstring{Kendall $\tau$ and its relationship with the Pearson correlation}
{Kendall tau and its relationship with the Pearson correlation}}\label{sec2.1}
Consider the random vectors $(X_i,Y_i), i=1,2,\ldots,n$, and the
Kendall $\tau$ rank correlation between $X_i$ and $Y_i$ is defined
as
%
\begin{equation}
\label{Kendall} {\tau}=\frac{1}{n(n-1)}\sum_{i\neq j}^{n}
\operatorname{sgn}(X_i-X_j)\operatorname{sgn}(Y_i-Y_j).
\end{equation}

Given this definition, it is easy to know that $|{\tau}|$ is
invariant against the monotonic transformation of $X_i$ or $Y_i$.
Furthermore, if $(X_i,Y_i)$ follows a~bivariate normal distribution
with mean zero and the Pearson correlation $\rho$, it can be shown
that [\citet{HubRon09}]
\[
\E(\tau) = \frac{2}{\pi} \arcsin\rho.
\]
In other words, when $(X_i,Y_i)$ follows bivariate normal
distribution, the Pearson correlation and Kendall $\tau$ have a
monotonic relationship in the following sense. If $|\rho|>c_1$ for a
given positive constant $c_1$, then there exists a positive constant
$c_2$ such that $|\E(\tau)|>c_2$, and if and only if $\rho=0$, $\E
(\tau) =0$. Such a~relationship helps us to obtain the sure
independence screening property for linear regression models under
the assumption of \citet{FanLv08} without any difficulties when
the Kendall $\tau$ is used.

When $(X_i, Y_i)$ are not bivariate normal but $\rho$ exists,
according to an approximation of the Kendall $\tau$ [\citet{Ken49}],
using the first fourth-order cumulants and the \textit{bivariate
Gram--Charlier series} expansion yield that
\begin{eqnarray*}
\label{Kendall-rho}
\E({\tau}) &\approx&\frac{2}{\pi}\arcsin(\rho)
\\
&&{}+\frac{1}{24\pi(1-\rho^2)^{3/2}} \bigl\{(\kappa_{40}+\kappa_{04})
\bigl(3\rho-2\rho^3\bigr)-4(\kappa_{31}+
\kappa_{13})+6\rho\kappa_{22} \bigr\},
\end{eqnarray*}
where $\kappa_{40}=\mu_{40}-3, \kappa_{31}=\mu_{31}-3\rho,
\kappa_{22}=\mu_{22}-2\rho^2-1$. If under some certain conditions
that $\kappa_{31}$ and $\kappa_{13}$ have a monotonic relationship
with $\rho$ and when $\rho=0$, $\kappa_{31}=0$ and $\kappa_{13}=0$,
intuitively
$\E(\tau)=0$ approximately when $\rho=0$, and if
\mbox{$|\rho|>c_1$}, then there may exist $c_2$ such that $|\E
(\tau)|>c_2$. This means that the Kendall' $\tau$ based method may
enjoy similar properties as the SIS enjoys without strong
conditions.

\subsection{Rank correlation screening}\label{sec2.2}
We start our procedure with the linear model as
%
\begin{equation}
\label{linearmodel}
\bfY=\bX\bbeta+\bvarepsilon,
\end{equation}
where $\bfY=(Y_1,\ldots,Y_n)^T$ is an
$n$-vector of response, $\bX=(\bfX_1,\ldots,\bfX_n)^T$ is an
$n\times p$ random design matrix with independent and identically
distributed $\bfX_1,\ldots, \bfX_n$,
$\bbeta=(\beta_1,\ldots,\beta_p)^T$ is a $p$-vector of parameters
and $\bvarepsilon=(\varepsilon_1,\ldots,\varepsilon_n)^T$ is an
$n$-vector of i.i.d. random errors independent of $\bX$.

To motivate our approach, we briefly review the SIS first. Let
%
\begin{equation}
\label{SIS-rank} \bomega=(\omega_1,\ldots,\omega_p)^T=
\bX^T\bfY,
\end{equation}
where each column of the $n\times p$ design matrix $\bX$ has been
standardized with mean zero and variance one. Then, for any given
$d_n < n$, take the selected submodel to be
\[
\widehat{\mathcal{M}}_{d_n}= \bigl\{ 1\le j \le p\dvtx |\omega_j|
\mbox{ is among the first } d_n \mbox{ largest of all} \bigr\}.
\]
This reduces the full model of size $p\gg n$ to a submodel with the
size $d_n$. By appropriately choosing $d_n$, all significant predictors
can be selected into the submodel indexed by
$\widehat{\mathcal{M}}_{d_n}$ with probability tending to 1; see
\citet{FanLv08}.

Similar to \citet{LiPenZhu11}, let
$\bomega=(\omega_1,\omega_2,\ldots,\omega_{p})^T$ be a $p$-vector
each being
%
\begin{equation}
\label{rankomega}\quad \omega_k=\frac{1}{n(n-1)}\sum
_{i\neq
j}^{n}I(X_{ik}<X_{jk})I(Y_i<Y_j)-
\frac{1}{4},\qquad k=1,\ldots,p,
\end{equation}
where $I(\cdot)$ denotes the indictor function, and $\omega_k$ is
the marginal rank correlation coefficient between $Y$ and
${\bfX}_{\cdot k}$, which is equal to a quarter of the Kendall $\tau$
between $Y$ and ${\bfX}_{\cdot k}$. As a U-statistic, $\omega_k$ is
easy to compute. We can\vspace*{1pt} then sort the magnitudes of all the
components of $\bomega=(\omega_1,\ldots,\omega_p)^T$ in a decreasing
order and select a submodel
%
\begin{equation}
\label{submodel1} \widehat{\mathcal{M}}_{d_{n}}=\bigl\{1\leq k\leq p\dvtx
|\omega_k|\mbox{ is among the first $d_n$ largest of all}\bigr\}
\end{equation}
or
%
\begin{equation}
\label{submodel}
\widehat{\mathcal{M}}_{\gamma_{n}}=\bigl\{1\leq k\leq p\dvtx |
\omega_k|> \gamma_n\bigr\},
\end{equation}
where $d_n$ or $\gamma_n$ is a predefined threshold value. Thus,
it shrinks the full model indexed $\{1,\ldots,p\}$ down to a
submodel indexed $\widehat{\mathcal{M}}_{d_{n}}$ or
$\widehat{\mathcal{M}}_{\gamma_{n}}$ with size
$|\widehat{\mathcal{M}}_{d_{n}}|<n$ or
$|\widehat{\mathcal{M}}_{\gamma_{n}}|<n$. Because of the
robustness of the Kendall $\tau$ against heavy-tailed distributions,
such a screening method is expected to be more robust than the SIS.

Consider a more general model as
%
\begin{equation}
\label{transmodel} H(Y_i)=\bfX_i^T
\bbeta+\varepsilon_i,\qquad i=1,\ldots,n,
\end{equation}
%
where $\varepsilon_i, i=1,\ldots,n$, are i.i.d. random errors
independent of $\bfX_i$ with mean zero and an unknown distribution $F$,
and $\bbeta=(\beta_1,\ldots,\beta_p)^T$ is a $p$-vector of parameters,
its norm constrained to 1 ($\|\bbeta\|=1$) for identifiability.
$H(\cdot)$ is an unspecified strictly increasing function. Model
(\ref{transmodel}) has been studied extensively in the econometric and
bioinformatic literature and is commonly used to stabilize the variance
of the error and to normalize/symmetrize the error distribution. With
different forms of $H$ and $F$, this model generates many different
parametric families of models. For example, when $H$ takes the form of
a power function and $F$ follows a normal distribution,
model~(\ref{transmodel}) reduces to the familiar Box--Cox transformation
models [\citet{BoxCox64,BicDok81}]. If $H(y)=y$ or $H(y)=\log(y)$,
model (\ref{transmodel}) reduces to the additive and multiplicative
error models, respectively. More parametric transformation models can
be found in the work of \citet{CarRup88}.

For model (\ref{transmodel}), the invariance against any strictly
increasing transformation yields that
%
\begin{eqnarray}
\label{rank} \omega_k &=&\frac{1}{n(n-1)}\sum
_{i\neq
j}^{n}I(X_{ik}<X_{jk})I(Y_i<Y_j)-
\frac{1}{4}
\nonumber\\[-8pt]\\[-8pt]
&=& \frac{1}{n(n-1)}\sum_{i\neq
j}^{n}I(X_{ik}<X_{jk})I
\bigl(H(Y_i)<H(Y_j)\bigr)-\frac{1}{4}\nonumber
\end{eqnarray}
for $k=1,\ldots,p$. That is, $\omega_k, k=1,2,\ldots,p$, can still
be applicable for the model with unknown transformation function.
Therefore, the RRCS method can also be applied to transformation
regression models that establish the nonlinear relationship between
the response and predictor variables.

\section{Sure screening properties of RRCS}\label{sec3}

In this section we study the sure screening properties of RRCS for
the linear regression model (\ref{linearmodel}) and the
transformation regression model (\ref{transmodel}). Without loss of
generality, let $(Y_{1}, X_{1k})$, $(Y_{2}, X_{2k})$ be the
independent copies of $(Y,X_{k})$, where $\E Y=\E X_k=0$ and $\E Y^2=\E
X_k^2=1, k=1,\ldots,p$, and assume that
\[
\mathcal{M}_{\ast}=\{1 \le k \le p\dvtx \beta_k \ne0\}
\]
is the true sparse model with nonsparsity size
$s_n=|\mathcal{M}_\ast|$, recalling that
$\bbeta=(\beta_1,\ldots,\beta_p)^T$ is the true parameter vector.
The compliment of $\mathcal{M}_{\ast}$ is
\[
\mathcal{M}_\ast^c=\{1 \le k \le p\dvtx k \notin
\mathcal{M}_\ast\}.
\]
Furthermore, for $k=1,\ldots, p$, let $\rho_k=\operatorname{corr}(X_k,Y)$
for model (\ref{linearmodel}) and
$\rho^\ast_k=\operatorname{corr}(X_k,H(Y))$ for model (\ref{transmodel}).
Recall the definition of $\bomega=\{\omega_1,\ldots,\omega_p\}^T$
in~(\ref{rankomega}) for both (\ref{linearmodel}) and
(\ref{transmodel}).

The following marginal conditions on the models are needed to ensure
the sure screening properties of
RRCS.\vspace*{8pt}

\textit{Marginally symmetric condition and Multi-modal condition}:\quad
For mod\-el~(\ref{linearmodel}):

\begin{longlist}[(M2)]
\item[(M1)]
Denote $\Delta Y=Y_{1}-Y_{2}$, then the conditional
distribution $F_{\Delta{Y}|\Delta{X}_{k}}(t)$ is symmetric about
zero when $k \in\mathcal{M}_\ast^c$, where $\Delta
X_{k}=X_{1k}-X_{2k}$.\vadjust{\goodbreak}

\item[(M2)] Denote $\Delta\epsilon_k=Y_{1}-Y_{2}-\rho_{k}(X_{1k}-X_{2k})$
and $\Delta X_{k}=X_{1k}-X_{2k}$, then the conditional
distribution $F_{\Delta\epsilon_{k}|\Delta{X}_{k}}(t)=\pi_{0k}
F_0(t,\sigma^2_0|\Delta{X}_{k})+(1-\pi_{0k})F_1(t,\break\sigma^2_1|\Delta{X}_{k})
$ follows a symmetric finite mixture distribution where
$F_0(t,\sigma_0^2|\Delta{X}_{k})$ follows a symmetric unimodal
distribution with the conditional variance $\sigma_0^2$ related to
$\Delta{X}_{k}$ and $F_1(t,\sigma_1^2|\Delta{X}_{k})$ is a symmetric
distribution function with the conditional variance $\sigma_1^2$
related to $\Delta{X}_{k}$ when $k \in\mathcal{M}_{*}$.
$\pi_{0k}\ge\pi^\ast$, where~$\pi^\ast$ is a given positive constant
in $(0,1]$ for any
$\Delta{X}_k$ and any $k \in\mathcal{M}_{*}$.
\end{longlist}

For model (\ref{transmodel}):

\begin{longlist}[(M2$'$)]
\item[(M1$'$)]
Denote $\Delta H(Y)=H(Y_{1})-H(Y_{2})$, where $H(\cdot)$ is
the link function of the transformation regression model
(\ref{transmodel}), and $\Delta X_{k}=X_{1k}-X_{2k}$. The
conditional distribution
$F_{\Delta
H(Y)|\Delta{X}_{k}}(t)$ is symmetric about zero when \mbox{$k \in\mathcal{M}_\ast^c$}.

\item[(M2$'$)] Denote $\Delta
\epsilon_k=H(Y_{1})-H(Y_{2})-\rho^\ast_{k}(X_{1k}-X_{2k})$ and $\Delta
X_{k}=X_{1k}-X_{2k}$, where $H(\cdot)$ is the link function of the
transformation regression mod\-el~(\ref{transmodel}), then the
conditional distribution
$F_{\Delta\epsilon_{k}|\Delta{X}_{k}}(t)=\pi_{0k}
F_0(t,\sigma^2_0|\Delta{X}_{k})+(1-\pi_{0k})F_1(t,\sigma^2_1|\Delta{X}_{k})
$ follows a symmetric finite mixture distribution where
$F_0(t,\sigma_0^2|\Delta{X}_{k})$ follows a symmetric unimodal
distribution with the conditional variance $\sigma_0^2$ related to
$\Delta{X}_{k}$ and $F_1(t,\sigma_1^2|\Delta{X}_{k})$ is a symmetric
distribution function with the conditional variance $\sigma_1^2$
related to $\Delta{X}_{k}$ when $k \in\mathcal{M}_{*}$.
$\pi_{0k}\ge\pi^\ast$, where $\pi^\ast$ is a given positive constant in
$(0,1]$ for any~$\Delta{X}_k$ and any $k \in\mathcal{M}_{*}$.
\end{longlist}

\begin{rema}\label{re1} According to the definition and
symmetric form of $\Delta Y, \Delta
X_k$ and~$\Delta\epsilon_k$, the marginally symmetric conditions
(M2) and (M2$'$) are very mild. When $\pi^\ast$ is small enough,
the distribution is close to $F_1$ which is naturally symmetric and has
no stringent constraint.

A special case is that the conditional
distribution of $\epsilon_{ik}=Y_i-\rho_k X_{ik}$ or
$\epsilon_{ik}=H(Y_i)-\rho^\ast_k X_{ik}$, given $X_{ik}$
$(i=1,\ldots,n)$, is homogeneous (not depending on $X_{ik}$) with a finite
number of modes.
Actually, when this condition holds, the conditional distribution of
$\epsilon_{ik}$
given $X_{ik}$ is identical to the corresponding unconditional marginal
distribution. Note that
$\Delta\epsilon_k=\epsilon_{1k}-\epsilon_{2k}$. When $\epsilon_{ik},
i=1,2$, follows multimodal distribution
$F_{\epsilon}(t)$ with no more than~$K$ modes where~$K$ is not
related to $k$ and $n$, such a distribution function can be
rewritten as a weighted sum of~$K$ unimodal distributions $F_i(\cdot)$ as
\[
F_{\epsilon}(t) = \sum_{i=1}^K
\pi_i F_i(t),
\]
where $\pi_i \ge0, i=1, \ldots,K$, with $\sum_{i=1}^K \pi_i=1$.
Then it is easy to see that the distribution of $\Delta
\epsilon_k=\epsilon_{1k}-\epsilon_{2k}$ has the following form:
\begin{eqnarray*}
F_{\Delta\epsilon}(t)&=& \sum_{i=1}^K\sum
_{j=1}^K \pi_i
\pi_j F^\ast_{ij}(t) =\sum
_{i=1}^K \pi_i^2
F^\ast_{ii}(t)+\sum_{i \ne j}^K
\pi_i\pi_j F^\ast_{ij}(t)
\\
&=& \Biggl\{\sum_{i=1}^K
\pi_i^2 \Biggr\} \sum_{i=1}^K
\frac{\pi_i^2}{\sum_{i=1}^K \pi_i^2} F^\ast_{ii}(t)+\Biggl(1-\sum
_{i=1}^K \pi_i^2\Biggr)\sum
_{i \ne
j}^K \frac{\pi_i\pi_j}{1-\sum_{i=1}^K \pi_i^2}F^\ast_{ij}(t)
\\
&\hat{=}& \pi_0^\ast F_0^{\ast\ast}(t) +
\bigl(1-\pi_0^\ast\bigr) F_1^{\ast\ast}(t),
\end{eqnarray*}
where $F^\ast_{ij}(t), i,j=1,\ldots,K$, are the distributions of the
differences of two independent variables, that is, $Z_i-Z_j$ where $Z_i$
follows the distribution of $F_i(t)$ and $Z_j$ follows the
distribution of $F_j(t)$, respectively. Because $F_i(t),
i=1,\ldots,K$, are unimodal distributions, $F_{ii}^\ast,
i=1,\ldots,K$, are then symmetric unimodal distributions. Hence,
$F_0^{\ast\ast}(t)$ is a symmetric unimodal distribution. It is also
easy to see that $F_1^{\ast\ast}(t)$ is a symmetric multimodal
distribution function. On the other hand, $\pi_0^\ast=\sum_{i=1}^K
\pi_i^2 \ge1/K (\sum_{i=1}^K \pi_i)^2=1/K$. As such, (M2) or (M2$'$) is
satisfied.
\end{rema}

Other than the marginally symmetric conditions, we also need the
following regularity conditions:

\begin{longlist}[(C1)]
\item[(C1)]
As $n \rightarrow+\infty$, the dimensionality of $\bfX$
satisfies $p=O(\exp(n^{\delta}))$ for some $\delta\in(0,1)$,
satisfying $\delta+2\kappa<1$ for any $\kappa\in(0,\frac{1}{2})$.

\item[(C2)]
$c_{\mathcal{M}_{*}}=\min_{k\in\mathcal{M}_{*}}\E|X_{1k}|$ is a
positive constant and is free of $p$.

\item[(C3)] The predictors $\bfX_i$ and the error $\varepsilon_i$,
$i=1,\ldots,n$, are independent of one another.
\end{longlist}

\begin{rema}\label{re2}
Condition (C1) guarantees that for the independence screening method,
we can select significant predictors into a working submodel with
probability tending to 1. SIS also needs this condition; see
\citet{FanLv08} and \citet{FanSon10}. Condition (C2) is a
mild technical condition that ensures the sure screening property of
the RRCS procedure. It is worth mentioning that we do not need to have
a uniform bound for all $\E X_{1k}^2$. If the size of $\mathcal{M}_{*}$
goes to infinity with a relatively slow speed, we can relax this
condition to $c_{\mathcal{M}_{*}}>cn^{-\iota}$ for some positive
constant $c$ and $\iota\in(0,1)$ with a suitable choice of the
threshold~$\gamma_{n}$. Precisely, $\gamma_{n}$ can be chosen as
$c'n^{-\kappa-\iota}$ for some positive constant $c'$ where $\kappa$
satisfies $2\kappa+2\iota<1$. From Theorem~\ref{theo1} below, we can
see that $|\E(\omega_{k})|>cn^{-\kappa-\iota}$ for $k\in
\mathcal{M}_{*} $. To ensure the sure screening properties, (C1) needs
to be changed to $\delta+2\kappa+2\iota<1$.
\end{rema}

\begin{theo}
\label{theo1} Under the regularity condition \textup{(C2)} and the
marginal symmetric conditions \textup{(M1)} and \textup{(M2)} for model
(\ref{linearmodel}), we have the following:

\begin{longlist}
\item
$\E(\omega_{k})=0$ if and only if $\rho_{k}=0$.
\item
If $|\rho_{k}|>c_{1}n^{-\kappa}$
for $k \in\mathcal{M}_{*}$ with a positive constant $c_1>0$,
then there exists a positive constant $c_2$ such that
$\min_{k \in
\mathcal{M}_{*}} |\E(\omega_{k})|>c_2n^{-\kappa}$.
\end{longlist}

For model (\ref{transmodel}), replacing conditions \textup{(M1)} and
\textup{(M2)} with~\textup{(M1$'$)} and~\textup{(M2$'$)}, then:

\begin{longlist}
\item[(i$'$)] $\E(\omega_{k})=0$ if and only if $\rho^\ast_{k}=0$.\vadjust{\goodbreak}
\item[(ii$'$)] If $|\rho^\ast_{k}|>c_{1}n^{-\kappa}$
for $k \in\mathcal{M}_{*}$ with a positive constant $c_1>0$,
then there exists a positive constant $c_2$ such that
$\min_{k \in
\mathcal{M}_{*}} |\E(\omega_{k})|>c_2n^{-\kappa}$.
\end{longlist}
\end{theo}

\begin{rema}\label{re3}
As \citet{FanSon10} mentioned, the marginally symmetric condition
(M1) is weaker than the partial orthogonality condition assumed by
\citet{HuaHorMa08}, that is, $\{X_{k}, k\in\mathcal{M}_{*}^{c}\}$
is independent of $\{X_{k}, k\in\mathcal{M}_{*}\}$, which can lead to
the model selection consistency for the linear model. Our results,
together with the following Theorem~\ref{theo2}, indicate that under
weaker conditions, consistency can also be achieved even for
transformation regression models. Furthermore, as in the discussion of
\citet{FanSon10}, a necessary condition for the sure screening is
that the significant predictors $X_k$ with $\beta_k \ne0$ are
correlated with the response in the sense that $\rho_k \ne0$. The
result (i) of Theorem~\ref{theo1} also shows that when the Kendall
$\tau$ is used, this property can be held, which suggests that the
insignificant predictors in $\mathcal{M}_{*}^{c}$ can be detected from
$\E(\omega_{k})$ at the population level. Result (ii) indicates that
under marginally symmetric conditions, a suitable threshold
$\gamma_{n}$ can entail the sure screening in the sense of
\[
\min_{k\in\mathcal{M}_{*}}\bigl|\E(\omega_{k})\bigr|\geq\gamma_{n},\qquad
\max_{k\in\mathcal{M}_{*}^{c}}\bigl|\E(\omega_{k})\bigr|=0.
\]
\end{rema}

\begin{rema}\label{re4}
As a by-product, Theorem~\ref{theo1} reveals the relationship
between the Pearson correlation and the Kendall $\tau$ under general
conditions, especially the multi-modal conditions (M2) or (M2$'$) which
in itself is of interest. However, either condition (M2) or (M2$'$) is
a sufficient condition to guarantee that the Kendall $\tau$ has
either the property (ii) or (ii$'$) of Theorem~\ref{theo1}, and then
has the sure screening property. As in the discussion in
Section~\ref{sec2.1}, following the high order
\textit{bivariate Gram--Charlier
series expansion} to approximate the joint distribution of
$(X_i,Y_i)$, under certain conditions such as either the
condition or sub-Gaussian tail condition, we could also obtain
similar results of Theorem~\ref{theo1}. It would involve some high
order of moments or cumulants. However, as shown in Theorem
\ref{theo1}, either the multi-modal condition (M2) or (M2$'$) is to
ensure the robust properties of the proposed RRCS, and depicts
those properties more clearly. Furthermore, we will show in the
proposition below that the bivariate normal copula family also makes
another sufficient condition for the following Theorem~\ref{theo1}
to hold.
\end{rema}



\textit{Bivariate normal copula family based marginal condition}:\quad
We give another sufficient condition for $(X_i, Y_i)$ for the
results of Theorem~\ref{theo1} to hold. Consider the bivariate
normal copula family which is defined as
\[
C_\theta(u_1,u_2)=\Phi_\theta\bigl(\Phi^{-1}(u_1),
\Phi^{-1}(u_2)\bigr),\qquad 0\le u_1,u_2 \le1,
\]
where $\Phi_\theta$ is a bivariate standard normal distribution
function with mean zero, variance one and correlation $\theta$,
$\Phi$ is the one-dimensional standard normal distribution function.
Let $\mathcal{F}$ denote the collection\vadjust{\goodbreak} of all distribution
functions on $\mathbb{R}$. We then define the bivariate distribution
family $\mathcal{P}$ as
\[
\mathcal{P}=\bigl\{ C_\theta\bigl(F_X(x),
F_Y(y)\bigr), (x,y) \in\mathbb{R}^2, F_X
\in\mathcal{F}, F_Y \in\mathcal{F}\bigr\}.
\]
Copula now is a popular tool to study the dependence among
multivariate random variables. For details, see \citet{Nel06}. The
normal copula family is an important copula family in practice.
Particularly, the bivariate normal copula family can be used to
approximate most of the distributions of bivariate continuous or
discrete random vectors, for example, see
\citet{CarNel,GhoHen03,PitChaKoh06} and \citet{ChaLEc09}.

Based on the results of \citet{KlaWel97} and the
monotonic relationship between the Kendall $\tau$ and the Pearson
correlation, the multi-modality can be replaced by the above copula
distribution family. A proposition is stated below.

\begin{prop}
\label{prop1}Under the marginal symmetric condition
\textup{(M1)} for mod\-el~(\ref{linearmodel}), we have the following:

\begin{longlist}
\item
$\E(\omega_{k})=0$ if and only if $\rho_{k}=0$.
\item
If $|\rho_{k}|>c_{1}n^{-\kappa}$ with a positive constant
$c_1>0$ and the joint distribution $F(x,y)$ of $(X_k,Y)$ is in
$\mathcal{P}$,
for $k \in\mathcal{M}_{*}$,
then there exists a positive constant $c_2$ such that
$\min_{k \in
\mathcal{M}_{*}} |\E(\omega_{k})|>c_2n^{-\kappa}$.
\end{longlist}

For model (\ref{transmodel}), replacing condition \textup{(M1)} with
\textup{(M1$'$)}, then:

\begin{longlist}[(ii$'$)]
\item[(i$'$)] $\E(\omega_{k})=0$ if and only if $\rho^\ast_{k}=0$.

\item[(ii$'$)] If $|\rho^\ast_{k}|>c_{1}n^{-\kappa}$ with a positive constant
$c_1>0$ and the joint distribution $F(x,y)$ of $(X_k,Y)$ is in
$\mathcal{P}$
for $k \in\mathcal{M}_{*}$,
then there exists a positive constant $c_2$ such that
$\min_{k \in
\mathcal{M}_{*}} |\E(\omega_{k})|>c_2n^{-\kappa}$.
\end{longlist}
\end{prop}

\begin{rema}\label{re5}
If the joint distribution of $(X,Y)$ is in $\mathcal{P}$ with
the formula $F(X,Y)=C_\theta(F_X(X), F_Y(Y))$, the results of
\citet{KlaWel97} suggested that $|\theta|$ equals the
maximum correlation coefficient between~$X$ and~$Y$. As shown in the
proof of the proposition, when we replace $\rho$ by $\theta$ in the
proposition, the results continue to hold. Hence, this proposition
provides a bridge between our method and the generalized correlation
proposed by \citet{HalMil09} because, according to their
definitions, the generalized correlation coefficient is an
approximation of the maximum correlation coefficient.
\end{rema}

\textit{Sure screening property of RRCS}:\quad
Based on Theorem~\ref{theo1} or Proposition~\ref{prop1}, the sure
screening property and model selection consistency of RRCS are
stated in the following results.

\begin{theo}
\label{theo2} Under the conditions \textup{(C1)--(C3)}, and the
conditions of Theorem~\ref{theo1} or Proposition~\ref{prop1}
corresponding to either model (\ref{linearmodel}) or
mod\-el~(\ref{transmodel}), for some\vadjust{\goodbreak} $0<\kappa<1/2$ and $c_{3}>0$, there exists
a positive constant $c_{4}>0$ such that
\[
{\mathbb{P} \Bigl(\max_{1 \le j \le p} \bigl|\omega_j-E(
\omega_j)\bigr|\ge c_3 n^{-\kappa} \Bigr) \le p \bigl\{
\exp\bigl(-c_4 n^{1-2\kappa}\bigr) \bigr\}.}
\]
Furthermore, by taking $\gamma_n =c_5n^{-\kappa}$ with $c_5 \le
c_2/2$, if $|\rho_k| > c_1 n^{-\kappa}$ for \mbox{$j \in
\mathcal{M}_\ast$}, we have
\[
\mathbb{P} (\mathcal{M}_\ast\subset\widehat{\mathcal{M}}_{\gamma_n}
) \ge1-2|\mathcal{M}_{*}|\bigl\{ \exp\bigl(-c_4
n^{1-2\kappa}\bigr)\bigr\}.
\]
\end{theo}

\begin{rema}\label{re6}
Theorem~\ref{theo2} shows that RRCS can handle the NP-dimension\-ality
problem for linear and semiparametric transformation regression models.
It also permits $\log p=o(n^{1-2\kappa})$, which is identical to that
in \citet{FanLv08} for the linear model and is faster than $\log
p=o(n^{(1-2\kappa)/A})$ with $A=\max(\alpha+4, 3\alpha+2)$ for some
positive $\alpha$ in \citet{FanSon10} when the likelihood ratio
screening is used.
\end{rema}

\begin{rema}\label{re7}
It is obvious when the joint distribution of $(\bfX^T_i,Y_i)$ follows
a multivariate normal distribution, conditions (M1) and (M2) are
automatically valid. The results of sure screening properties are
equivalent to those of \citet{FanLv08} under weaker conditions. This
is because of the definition of the rank correlation Kendall $\tau$ and
its monotonic relationship with the Pearson correlation as in the
discussion in Section~\ref{sec2}. The Kendall $\tau$ can be regarded as
a U-statistic and uses the indicator function as the link function. As
the indicator function is a bounded function, the exponential
U-statistic inequality can be used to directly control the tail of the
rank correlation Kendall $\tau$ rather than those of $\bfX_i$ and
$Y_i$.

Under the conditions of Proposition~\ref{prop1}, following similar
steps, the same results of Theorem~\ref{theo2} and the following Theorem~\ref{theo3} can
be obtained without any difficulties. Thus, we only present the
relevant results without the detailed technical proofs.
\end{rema}

The following theorem states that the size of
$\widehat{\mathcal{M}}_{\gamma_{n}}$ can be controlled by the RRCS
procedure.

\begin{theo}
\label{theo3}Under the conditions \textup{(C1)--(C3)}, and conditions
of Theorem~\ref{theo1} or Proposition~\ref{prop1} for model
(\ref{linearmodel}), when $|\rho_{k}|>c_{1}n^{-\kappa}$ for some
positive constant $c_{1}$ uniformly in $k \in\mathcal{M}_{*}$, for
any $\gamma_n= c_5 n^{-\kappa}$ there exists a constant $c_6>0$ such
that
%
\begin{equation}
\label{res-theo3}
\mathbb{P} \bigl(|\widehat{\mathcal{M}}_{\gamma_n}|
\le O \bigl\{ n^{2\kappa} \lambda_{\max}(\Sigma)\bigr\} \bigr) \ge 1-p
\bigl\{\exp\bigl(-c_6 n^{1-2\kappa}\bigr) \bigr\},
\end{equation}
where $\Sigma=\operatorname{Cov}({\bfX}_i)$ and
$\mathbf{\bfX}_i=(X_{i1},\ldots,X_{ip})$. For model
(\ref{transmodel}) in addition to conditions \textup{(C1)--(C3)} and the
marginal symmetric conditions \textup{(M1$'$)} and \textup{(M2$'$)}, when
$|\rho^\ast_{k}|>c_{1}n^{-\kappa}$ for some positive constant
$c_{1}$ uniformly in $k \in\mathcal{M}_{*}$ and
$\operatorname{Var}(H(Y))=O(1)$, for $\gamma_n= c_5 n^{-\kappa}$ there
exists a constant $c_6>0$ such that the above inequality
(\ref{res-theo3}) holds.
\end{theo}

\begin{rema}\label{re8}
Compared with Theorem 5 of \citet{FanSon10}, the conditions of
Theorem~\ref{theo3} are much weaker and the obtained inequalities are
much simpler in form although the rates are similar. The number of
selected predictors is of the order $\|\Sigma\bbeta\|/\gamma_n^2$,
which is bounded by $O\{n^{2\kappa}\lambda_{\max}(\Sigma)\}$ when
$\operatorname{Var}(H(Y))=O(1)$. Hence, when
$\lambda_{\max}(\Sigma)=O(n^\tau)$, the size of the selected predictors
is of the order $O(n^{2\kappa+\tau})$, which can be smaller than $n$
when $2\kappa+\tau<1$.
\end{rema}

From Theorems~\ref{theo1}--\ref{theo3}, the rank correlation
has sure screening properties and model selection consistency.
However, it is also obvious that it does not sufficiently use all of
the information from data, particularly the correlations of predictors.
Hence, as most of the other sure screening methods, the rank sure
screening can be only regarded as an initial model selection
reducing the ultra-high dimension down to a dimension smaller than
the sample size $n$ without losing any important significant predictor
variables. As the numerical results in Section~\ref{sec5} and
the discussion of \citet{FanLv08} show, the correlation of predictors
could seriously affect the sure screening results, and thus more
subtle sure screening methods, such as Iterative Sure Independence
Screening (ISIS) [\citet{FanLv08}], are in need.


\section{IRRCS: Iterative robust rank correlation screening}\label{sec4}

\subsection{IRRCS}\label{sec4.1}

With RRCS, the dimension can be brought down to a value smaller than
the sample size with a probability tending to one.
Thus, we can work
on a smaller submodel. However, 
in most situations, RRCS can be only regarded as a crude model
selection method, and the resulting model may still contain many
superfluous predictors. It is partly because strong correlation
always exists between predictors when too many predictors are
involved [see \citet{FanLv08}], and the basic sure screening
methods do not use this correlation information. We also face some
other issues. First, in modeling high dimensional data, it is often
a challenge to determine outliers. High dimensionality also
increases the likelihood of extreme values of predictors. Second,
even when the model dimension is smaller than the sample size, the
design matrix may still be near singular when strong correlation
exists between predictors. Third, the usual normal or sub-Gaussian
distributional assumption on predictors/errors is not easy to
substantiate. Fourth, it is also an unfortunate fact that the RRCS
procedure may break down if a~predictor is
marginally unrelated but jointly related with the response, or if a
predictor is jointly unrelated with the response
but has higher marginal correlation with the response than some
significant predictors. To deal with these issues, we develop a
robust iterative RRCS (IRRCS) that is motivated by the concept of
Iterative Sure Independence Screening (ISIS) in \citet{FanLv08}.

To this end, we first briefly describe a penalized smoothing
maximum rank correlation estimator (PSMRC) suggested by \citet{LinPen}. This estimation approach is applied to simultaneously
further select and estimate a final working submodel through working
on $\bbeta$.

For model (\ref{transmodel}), the monotonicity of $H$ and the
independence of $\bX$ and~$\bvarepsilon$ ensure that
\[
\mathbb{P}(Y_i\geq Y_j|\bfX_i,
\bfX_j)\geq\mathbb{P}(Y_i\leq Y_j|
\bfX_i,\bfX_j) \qquad\mbox{whenever } \bfX_i^T
\bbeta\geq\bfX_j^T\bbeta.
\]
Hence, $\bbeta$ can be estimated by maximizing
%
\begin{equation}
G_n(\bbeta)=\frac{1}{n(n-1)}\sum_{i\neq
j}I
(Y_i>Y_j )I \bigl({\bfX}_i^T
\bbeta>\bfX_j^T \bbeta\bigr).
\end{equation}
It is easy to see that $G_n(\bbeta)$ is another version of the
Kendall $\tau$ between $Y_i$ and $\bfX_i^T\bbeta$. The maximum rank
correlation [MRC; \citet{Han87}] estimator~$\hat{\bbeta}_n$ can be
applied to estimate $\bbeta$. When $p$ is fixed, 
the $n^{1/2}$-consistency and the asymptotic normality of
$\hat{\bbeta}_n$ have been derived. However, because $G_n(\bbeta)$
is not a smooth function, the Newton--Raphson algorithm cannot be
used directly, and the optimization of $G_n(\bbeta)$ requires an
intensive search at heavy computational cost. We then consider
PSMRC as follows. Define
%
\begin{equation}
\label{penaQ} L_n(\bbeta)=S_n(\bbeta)-\sum
_{j=1}^d p_{\lambda_n}\bigl(|
\beta_j|\bigr)
\end{equation}
and
%
\begin{equation}
S_n(\bbeta)= \frac{1}{n(n-1)}\sum_{i\neq
j}I
(Y_i>Y_j )\Phi\bigl((\bfX_i-
\bfX_j)^T \bbeta/h \bigr),
\end{equation}
where $\Phi(\cdot)$ is the standard normal distribution function, a
smooth function for the purpose of reducing computational burden,
$h$ is a small positive constant, and $p_{\lambda}(|\cdot|)$ is a
penalty function of $L_1$ type such as that in LASSO, SCAD or MCP. It
is easy to
see if $h\to0$, $\Phi((\bfX_i-\bfX_j)^T \bbeta/h ) \to
I ({\bfX}_i^T \bbeta>\bfX_j^T \bbeta)$. As $L_n(\bbeta)$
is a smoothing function of $\bbeta$, traditional optimal methods,
such as the Newton Raphson algorithm or newly developed LARS
[\citet{Efretal04}] and LLA [\citet{ZouLi08}], can be used to
obtain the maximizer of $L_n(\bbeta)$ to simultaneously achieve the
selection and estimation of $\bbeta$. For model
(\ref{linearmodel}), the problem is easier and we do not repeatedly
describe the estimation for it.

Next, we introduce our intuitive idea for the proposed IRRCS for the
transformation regression model. Such an idea can be also applied to
the linear model since it is a special transformation regression
model. In fact, given the i.i.d. sequences $Y_i$ and $\bfX_i^T
\bbeta, i=1,\ldots,n$, define $Y_{ij}^\ast=I(Y_i<Y_j)$ and
$\bfX^\ast_{ij}(\bbeta)=I(\bfX_{i}\bbeta<\bfX_{j}\bbeta)$. Then the
Pearson correlation between $Y^\ast_{ij}$ and
$\bfX^\ast_{ij}(\bbeta)$ is the rank correlation Kendall $\tau$
between $Y_i$ and $\bfX_{i}\bbeta$. According to the idea of the
maximum rank correlation [MRC; \citet{Han87}] estimator, the estimate of
$\bbeta$ for the transformation regression model just maximizes the
Pearson correlation between $Y^\ast_{ij}$ and
$\bfX^\ast_{ij}(\bbeta)$ or the rank correlation Kendall $\tau$
between $Y_i$ and $\bfX_{i}\bbeta$. If we do not care about the norm
of $\bbeta$, the least squares estimate of $\bbeta$ in the linear
model just maximizes the Pearson correlation between $Y_i$ and
$\bfX_i^T \bbeta$. If we regard the transformation model as the
following special linear model:
\[
Y^\ast_{ij}=\bfX^\ast_{ij}(\bbeta)+
\varepsilon_{ij},
\]
where $\varepsilon_{ij}=I(\varepsilon_i<\varepsilon_j)$. Then it is
easy to see that MRC for the transformation model and the least
squares estimate for the linear model are based on a similar
principle and, hence, the idea of Iterative Sure Independence
Screening (ISIS) for the linear model in \citet{FanLv08} can be
used for the transformation model. Based on this intuitive insight,
our proposed IRRCS procedure is as follows:

\begin{longlist}[\textit{Step} 1.]
\item[\textit{Step} 1.] First the RRCS procedure is used to reduce the original
dimension to a value $[n/\log n]$ smaller than $n$. Then, based on the
joint information from the $[n/\log n]$ predictors that survive after
the RRCS, we select a subset of $d_1$ predictors
$\mathcal{M}_1=\{X_{i_1},\ldots,X_{i_{d_1}}\}$ by a model selection
method such as the nonconcave penalized M-estimation proposed by
\citet{LiPenZhu11} for model (\ref{linearmodel}) and the penalized
smoothing maximum correlation estimator [\citet{LinPen}] for model
(\ref{transmodel}).
\item[\textit{Step} 2.] Let $\bfX_{i,\mathcal{M}_1}=
(X_{i_1},\ldots,X_{i_{d_1}})^T$ be the $d_1\times1$ vector selected in
step~1, and
$l=1,\ldots,p-d_1$.
\begin{itemize}
\item For model (\ref{linearmodel}), define $Y_i^\ast=Y_i-\bfX
_{i,\mathcal{M}_1}^{T}\hat{\bbeta}_{\mathcal{M}_1}$,
then the Kendall $\tau$ values for the remaining $p-d_1$ predictors
are calculated as follows:
\[
\omega_l=\frac{1}{n(n-1)}\sum_{j\neq i}^n
I \bigl(Y^\ast_i<Y_j^\ast\bigr)
I(X_{il}<X_{jl})-\frac{1}{4},
\]
where $\hat{\beta}_{\mathcal{M}_1}$ is a vector estimator of the
$d_1$ nonzero coefficients that are estimated by the nonconcave
penalized M-estimate method in \citet{LiPenZhu11}. Sort the
$p-d_1$ values of the $|\omega_l|$ again and select another subset
of $[n/\log n]$ predictors from $\mathcal{M}-\mathcal{M}_1$.

\item For model (\ref{transmodel}), define
$I(Y^\ast_i,Y^\ast_j)=I(Y_i,Y_j)-I(\bfX_{i,\mathcal{M}_1}^{T}\hat
{\bbeta}_{\mathcal{M}_1}<\bfX_{j,\mathcal{M}_1}^{T}\hat{\bbeta
}_{\mathcal{M}_1})$
where $I(Y_i,Y_j)=I(Y_i<Y_j)$ where $\hat{\bbeta}_{\mathcal{M}_1}$
is an estimator of the $d_1$ nonzero coefficients, which are
estimated with the penalized smoothing maximum correlation estimator
of \citet{LinPen}. Then, compute the Kendall~$\tau$ through the
remaining $p-d_1$ predictors as
\[
\omega_l=\frac{1}{n(n-1)}\sum_{j\neq i}^n
I\bigl(Y^\ast_i,Y^\ast_j\bigr)
I(X_{il}<X_{jl})-\frac{1}{4},
\]
and sort the $p-d_1$ values of the $|\omega_l|$'s again and select
a subset of $[n/\log n]$ predictors as in step 1.

\end{itemize}
\item[\textit{Step} 3.] Replace\vspace*{1pt} $Y_i$ by $Y_i^\ast$ in
(\ref{linearmodel}) and $I(Y_i,Y_j)$ with $I(Y_i^\ast,Y_j^\ast)$ in
(\ref{penaQ}), and select a subset of $d_2$ predictors
$\mathcal{M}_2=\{X_{i_1},\ldots,X_{i_{d_2}}\}$ from the\vspace*{2pt}
joint information of the $[n/\log n]$ predictors that survived in step
2 as in step 1.

\item[\textit{Step} 4.] Iterate steps 2 and 3 until $k$ disjoint subsets $\mathcal
{M}_1,\ldots,\mathcal{M}_k$ are obtained
whose union $\mathcal{M}=\bigcup_{i=1}^{k}\mathcal{M}_i$ has a size $d$
less than sample size $n$. In the implementation, we can choose, for
example, the largest $k$ such that $|\mathcal{M}|<n$.
\end{longlist}

\subsection{Discussion on RRCS for generalized linear and single-index models}\label{sec4.2}

Consider the generalized linear model
%
\begin{equation}
\label{genlinearmodel} f_Y(y,\theta)=
\exp\bigl\{y\theta-b(\theta)+c(y)\bigr\}
\end{equation}
for known functions $b(\cdot)$ and $c(\cdot)$ and unknown function
$\theta$, where the dispersion parameter is not considered as the
mean regression modeled. The function $\theta$ is usually called
canonical or a natural parameter, and the following structure of
the generalized linear model is often considered:
%
\begin{equation}
\label{genmean} \E(Y|\bfX=\mathbf{x})=b'
\bigl(\theta(\mathbf{x})\bigr)=g^{-1} \Biggl(\sum
_{j=0}^p \beta_j x_j
\Biggr),
\end{equation}
where $\mathbf{x}=(x_0,\ldots,x_p)^T$ is a
$(p+1)$-dimensional predictor, $x_0=1$ represents the intercept, and
$\theta(\mathbf{x})=\sum_{j=0}^p \beta_j x_j$. In
this case,\vspace*{-1pt}
$g(\cdot)$ should be a strictly increasing function. Thus, we may
use $\omega$ of (\ref{rank}) with function $g^{-1}$ to rank the
importance of the predictors. Although the idea seems
straightforward, the technical details are not easily handled, and
we leave them to further study. In the simulations, we examine its
performance; see the details in Section~\ref{sec5}. In addition,
after reducing the dimension, we consider estimating the parameters
in the working submodel. Again, we can also see that
\[
\mathbb{P}(Y_i\geq Y_j|\bfX_i,
\bfX_j)\geq\mathbb{P}(Y_i\leq Y_j|
\bfX_i,\bfX_j) \qquad\mbox{whenever } \bfX_i^T
\bbeta\geq\bfX_j^T\bbeta.
\]
Hence, Han's (\citeyear{Han87}) MRC estimator can be used.
\citet{FanSon10} applied the idea of SIS to (\ref{genlinearmodel})
with NP-dimensionality, and used the maximum marginal likelihood
estimator (MMLE). They showed that the MMLE $\beta_j^M=0$ if and only
if
$\operatorname{Cov}(b'(\bfX^T\bbeta),X_j)=\operatorname{Cov}(Y,X_j)=0$.
That is, MMLE is equivalent to the Pearson correlation in a certain
sense when SIS is applied.

A further generalization is with unknown canonical link function
$g(\cdot)$. In this case, the generalized linear model can be
regarded as a special single index model with a strictly increasing
restriction as the link function $b'(\cdot)$ or $g(\cdot)$. Based on
the discussion in Section~\ref{sec2}, we can also use the Kendall
$\tau$ based method to select predictors and PSMRC to estimate the
parameters. The selection and estimation could be more robust than
with the MMLE based SIS.

\section{Numerical studies and application}\label{sec5}

\subsection{Simulations}\label{sec5.1}

In the first 4 examples, we compare the performance of the five
methods: SIS, ISIS, RRCS, IRRCS, and the generalized correlation rank
method (gcorr) proposed by \citet{HalMil09} by computing the
frequencies with which the selected models include all of the variables
in the true model, that is, their ability to correctly screen
unimportant variables. The simulation examples cover the linear models
used by \citet{FanLv08}, the transformation models used by
\citet{LinPen}, the Box--Cox transformation model used by
\citet{HalMil09}, and the generalized linear models used by
\citet{FanSon10}. We also use a ``semi-real'' example as Example
\ref{Example5}, in which a part of the data are from a real data set
and the other part of the data are artificial. The difference from the
other examples is that this data set contains categorical data.

\begin{Exam}\label{Example1}
Consider the following linear model:
%
\begin{equation}
\label{LM} Y_i=\bfX_i^{T}\bbeta+
\varepsilon_i,\qquad i=1,\ldots, n,
\end{equation}
%
where $\bbeta=(5,5,5,0,\ldots,0)^{T}$, $\bfX_i=
(X_{1i},\ldots,X_{pi})^T$ is a $p$-dimensional predictor and the noise
$\varepsilon_i$ is independent
of the predictors, and is generated from three different
distributions: the standard normal, the standard normal with 10\%
of the outliers following the Cauchy distribution and the standard $t$
distribution with three degrees of freedom. The first $k=3$
predictors are significant, but the others
are not.
$\bfX_i$ are generated from a multivariate normal distribution
$N(0,\Sigma)$ with entries of $\Sigma=(\sigma_{ij})_{p\times p}$
being $\sigma_{ii}=1,i=1,\ldots,p$, and $\sigma_{ij}=\rho,i\neq j$.
For some combinations with $p=100,1000$, $n=20,50,70$ and
$\rho=0,0.1,0.5,0.9$, the experiment is repeated 200 times.

As different methods may select a working model with different sizes,
to ensure a fair comparison, we select the same size of $n-1$
predictors using the four methods. Then we check their selection
accuracy in including the true model
$\{X_1,X_2,X_3\}$. The details of ISIS can be found in Section 4 of
\citet{FanLv08}. 
In Table~\ref{tab1}, we report the proportions of predictors
containing the true model selected by RRCS, SIS, IRRCS
and ISIS. 

From Table~\ref{tab1}, we can draw the following conclusions:

\begin{longlist}[(2)]
\item[(1)]
When noise $\varepsilon$ is drawn from the standard normal,
SIS and ISIS perform better than RRCS and IRRCS according to higher
proportions of predictors containing the true model selected.
The difference becomes smaller with a larger sample size and smaller
$\rho$. ISIS and IRRCS can greatly improve the
performance of SIS and RRCS. IRRCS can outperform ISIS. 

\begin{sidewaystable}
\tablewidth=\textheight
\tablewidth=\textwidth
\caption{Example \protect\ref{Example1}: the proportion of
predictors containing the true model $\{X_{1},X_{2},X_{3}\}$
selected by RRCS, SIS, IRRCS and ISIS}\label{tab1}
\begin{tabular*}{\tablewidth}{@{\extracolsep{\fill}}lcd{1.3}d{1.3}d{1.3}d{1.3}
d{1.3}d{1.3}d{1.3}d{1.3}d{1.3}d{1.3}d{1.3}d{1.3}@{}}
\hline
& \multicolumn{1}{c}{$\bolds{\varepsilon\sim}$} &
\multicolumn{4}{c}{$\bolds{N(0,1)}$} & \multicolumn{4}{c}{$\bolds{N(0,1)}$
\textbf{with 10\% outliers}} & \multicolumn{4}{c@{}}{$\bolds{t(3)}$}\\[-4pt]
& \multicolumn{1}{c}{\hrulefill} &
\multicolumn{4}{c}{\hrulefill} & \multicolumn{4}{c}{\hrulefill}
& \multicolumn{4}{c@{}}{\hrulefill}\\
\multicolumn{1}{@{}l}{$\bolds{(p,n)}$}
&\multicolumn{1}{c}{\textbf{Method}}
& \multicolumn{1}{c}{$\bolds{\rho=0}$}
& \multicolumn{1}{c}{$\bolds{0.1}$} & \multicolumn{1}{c}{$\bolds{0.5}$}
& \multicolumn{1}{c}{$\bolds{0.9}$} & \multicolumn{1}{c}{$\bolds{0}$}
& \multicolumn{1}{c}{$\bolds{0.1}$} & \multicolumn{1}{c}{$\bolds{0.5}$}
& \multicolumn{1}{c}{$\bolds{0.9}$} & \multicolumn{1}{c}{$\bolds{0}$}
& \multicolumn{1}{c}{$\bolds{0.1}$} & \multicolumn{1}{c}{$\bolds{0.5}$} &
\multicolumn{1}{c@{}}{$\bolds{0.9}$}\\
\hline
$(100,20)$ &RRCS & 0.765 & 0.745 & 0.605 & 0.405 & 0.840 &0.835 & 0.730 &
0.640 &
0.850 & 0.840 & 0.765 & 0.520 \\
& SIS & 0.835 & 0.875 & 0.725 & 0.650 & 0.810 & 0.845 & 0.705 & 0.590 &
0.775 &
0.805 & 0.600 & 0.315 \\
& IRRCS & 0.840 & 0.905 & 0.865 & 0.915 & 0.995 & 0.980 & 0.960 & 0.895
& 0.995
& 1 & 0.995 & 0.930 \\
& ISIS & 1 & 1 & 0.985 & 0.985 & 0.885 & 0.850 & 0.855 & 0.845 & 0.895
& 0.910
& 0.865 & 0.845 \\
[4pt]
$(100,50)$ &RRCS & 1 & 1 & 1 & 0.985 & 0.980 &0.960 &0.970 &0.930 & 1 &
0.995 &
0.980 & 0.965 \\
& SIS & 1 & 1 & 1 & 1 & 0.960 &0.950 &0.970 &0.915 & 0.965 & 0.970 &
0.960 &
0.920 \\
& IRRCS & 1 & 1 & 1 & 1 &1 & 1 & 1 &0.970 & 1 & 1 & 1 & 0.990 \\
& ISIS & 1 & 1 & 1 & 1 &0.985 &0.975 &0.975
&0.945 & 1 & 1 & 0.980 & 0.955\\
[8pt]
$(1000,20)$&RRCS & 0.145 & 0.165 & 0.060 & 0.235 & 0.245 & 0.250 & 0.155
& 0.110
& 0.245 & 0.325 & 0.225 & 0.150 \\
& SIS & 0.255 & 0.285 & 0.110 & 0.140 &0.250 & 0.265 & 0.125 & 0.110 &
0.300 &
0.270 & 0.220 & 0.110 \\
& IRRCS & 0.475 & 0.460 & 0.480 & 0.345 & 0.825 & 0.840 &0.620 & 0.465
& 0.860 &
0.895 & 0.680 & 0.580 \\
& ISIS & 0.835 & 0.865 & 0.715 & 0.530 & 0.795 & 0.840 & 0.650 &0.430 &
0.805 &
0.855 & 0.630 & 0.460 \\
[4pt]
$(1000,50)$&RRCS & 0.990 & 0.970 & 0.825 & 0.570 & 0.945 & 0.990 & 0.755
&0.555 &
1 & 0.990 & 0.930 & 0.750 \\
& SIS & 1 & 0.985 & 0.935 & 0.835 & 0.950 &0.985 & 0.845 & 0.655 &
0.985 & 0.985
& 0.810 & 0.620 \\
& IRRCS & 1 & 1 & 0.990 & 0.995 & 0.980 &0.995 &0.950 &0.865 & 1 & 1 &
1 &
0.985 \\
& ISIS & 1 & 1 & 1 & 0.995 & 0.955 & 0.990 &0.940 &0.850 & 1 & 0.990 &
0.935 &
0.850 \\
[4pt]
$(1000,70)$&RRCS & 1 & 1 &0.990 &0.870 & 0.945 & 0.990 & 0.965 & 0.835 &
1 & 1
& 0.980 & 0.860 \\
& SIS & 1 & 1 & 0.990 & 0.965 & 0.960 &0.950 & 0.925 & 0.875 & 1 &
0.990 &
0.950 & 0.850 \\
& IRRCS & 1 & 1 & 1 & 1 & 1 & 1 & 0.975 &0.965 & 1 & 1 & 1 & 1 \\
& ISIS & 1 & 1 & 1 & 1 & 0.970 & 0.960 & 0.950 &0.940 & 1 & 1 & 0.980 &
0.960\\
\hline
\end{tabular*}
\end{sidewaystable}

\item[(2)] When $\rho=0.5$ or 0.9, SIS and RRCS perform worse than in
the cases with $\rho=0$ or 0.1. This coincides with our intuition
that high collinearity deteriorates the performance of SIS and RRCS.

\item[(3)] It is also worth mentioning that even when there are outliers
or the heavy-tailed errors, RRCS is\vadjust{\goodbreak} not necessarily better than SIS.
This is an interesting observation. However, when we note the
signal-to-noise ratio, we may have an answer. Regardless of
outliers, model (\ref{LM}) has a large signal-to-noise ratio by
taking the nonzero coefficients $(\beta_1,\beta_2,\beta_3)=(5,5,5)$.
This means that the impact of the outliers on the results is
relatively small and RRCS, a nonparametric method, may not be able
to show its advantages. We have also tried other simulations with
smaller signal-to-noise ratios or larger percentages of outliers.
When data has larger percentages of outliers, the performance of
RRCS was better than SIS. Especially when iteration is used, IRRCS
can outperform the corresponding ISIS even in the case without
outliers. When the data has smaller signal-to-noise ratios, for
example, $(\beta_1,\beta_2,\beta_3, 0, \ldots, 0)=(1,2/3,1/3,
0,\ldots, 0)$, though the performance of SIS and RRCS are comparable
and encouraging, all of the results are not as good as the results
of SIS and RRCS in Table~\ref{tab1}. This is reasonable, as for all
variable selection methods, the phenomenon is the same: when
the signal-to-noise ratio becomes smaller, selecting significant
predictors gets more difficult.

\item[(4)] When the data are contaminated with 10\% outliers or are
generated from the $t(3)$ distribution, the IRRCS performs better
than the
ISIS procedure because we use the nonconcave penalized M-estimation in
the iterative step for IRRCS. 
\end{longlist}
\end{Exam}

\begin{Exam}\label{Example2}
Consider Example III in Section 4.2.3 of
\citet{FanLv08} with the underlying model, for $\bfX=(X_1, \ldots,
X_p)^T$,
%
\begin{equation}
\label{example3} Y=5X_1+5X_2+5X_3-15\sqrt{
\rho}X_4+X_5+\varepsilon,
\end{equation}
except that $X_1, X_2, X_3$ and noise $\varepsilon$ are distributed
identical to those in Example~\ref{Example1} above. For model
(\ref{example3}), $X_4\sim N(0,1)$ has correlation coefficient
$\sqrt{\rho}$ with all other $p-1$ variables, whereas $X_5\sim N(0,1)$
is uncorrelated with all the other $p-1$ variables. $X_5$ has the same
proportion of contributions to the response as $\varepsilon$ does, and
has an even weaker marginal correlation with $Y$ than $X_6,\ldots,X_p$
do. We take $\rho=0.5$ for simplicity. We generate 200 data sets for
this model and report in Table~\ref{tab3} the proportion of RRCS, SIS,
IRRCS and ISIS that can include the true model.

\begin{table}
\tabcolsep=0pt
\caption{For Example \protect\ref{Example2}: the proportion
of RRCS, SIS, IRRCS and ISIS that include the true model
$\{X_{1},X_{2},X_{3},X_{4},X_5\}$ $(\rho=0.5)$}
\label{tab3}
{\fontsize{8.7pt}{11pt}\selectfont{
\begin{tabular*}{\tablewidth}{@{\extracolsep{4in minus 4in}}lccd{1.3}d{1.3}c
d{1.3}d{1.3}cd{1.3}d{1.3}@{}}
\hline
& \multicolumn{1}{c}{$\bolds{\varepsilon\sim}$} &
\multicolumn{3}{c}{$\bolds{N(0,1)}$} & \multicolumn{3}{c@{}}{\hspace*{-1pt}$\bolds{N(0,1)}$
\textbf{with 10\% outliers}} & \multicolumn{3}{c@{}}{$\bolds{t(3)}$}\\[-4pt]
& \multicolumn{1}{c}{\hrulefill} &
\multicolumn{3}{c}{\hrulefill} & \multicolumn{3}{c}{\hspace*{-2pt}\hrulefill\hspace*{-2pt}}
& \multicolumn{3}{c@{}}{\hrulefill}\\
$\bolds{p}$& \textbf{Method} & \multicolumn{1}{c}{$\bolds{n=20}$}
& \multicolumn{1}{c}{$\bolds{n=50}$} & \multicolumn{1}{c}{$\bolds{n=70}$}
& \multicolumn{1}{c}{$\bolds{n=20}$} & \multicolumn{1}{c}{$\bolds{n=50}$}
& \multicolumn{1}{c}{$\bolds{n=70}$} & \multicolumn{1}{c}{$\bolds{n=20}$}
& \multicolumn{1}{c}{$\bolds{n=50}$} & \multicolumn{1}{c@{}}{$\bolds{n=70}$}
\\
\hline
\hphantom{0}100 &RRCS & 0 & 0.305 &0.595 & 0 &0.220 &0.575 & 0 &0.305 &0.575 \\
& SIS & 0 &0.285 &0.535 & 0 & 0.195 & 0.525 & 0 &0.240 &0.535 \\
& IRRCS & 0 &0.500 &0.820 & 0 &0.495 &0.815 & 0 &0.530 &0.805 \\
& ISIS & 0 &0.465 &0.855 & 0 &0.415 &0.805 & 0 &0.405 &0.775 \\
1000 &RRCS & 0 & 0 & 0 & 0 &0 &0 & 0 & 0 & 0 \\
& SIS & 0 & 0 & 0 & 0 & 0 & 0 & 0 & 0 & 0 \\
& IRRCS & 0 &0.035 &0.085 & 0 &0.030 & 0.055 & 0 &0.030 &0.085 \\
& ISIS & 0 &0.045 &0.090 & 0 &0.015 &0.035 & 0 & 0 &0.020 \\
\hline
\end{tabular*}}}
\end{table}

The results in Table~\ref{tab3} allow us to draw different
conclusions than those from Example~\ref{Example1}. Even in the case without
outliers or the heavy-tailed errors, SIS and ISIS are not
definitely better than RRCS and IRRCS, respectively, whereas in the
cases with outliers or heavy-tailed errors there is no exception for
IRRCS to work well and better than ISIS. However, the small
proportions of RRCS and SIS show their bad performance.
\end{Exam}

\begin{Exam}\label{Example3}
Consider the following generalized
Box--Cox transformation model:
%
\begin{equation}
\label{box-coxmodel} H(Y_i)=\bfX_i^{T}
\bbeta+\varepsilon_i,\qquad i=1,2,\ldots,n,
\end{equation}
where the transformation functions are unknown. In the simulations,
we consider the following forms:
\begin{itemize}
\item Box--Cox transformation, $\frac{|Y|^{\lambda}
\operatorname{sgn}(Y)-1}{\lambda}$, where
$\lambda=0.25,0.5,0.75$;\vspace*{1pt}
\item Logarithm transformation function, $H(Y)=\log Y$.
\end{itemize}

The linear regression model and the logarithm transformation model
are special cases of the generalized Box--Cox transformation model
with $\lambda= 1$ and $\lambda= 0$, respectively. Again, noise
$\varepsilon_i$ follows the distributions as those in the above
examples, $\bbeta= (3, 1.5,2, 0,\ldots,0)^T$ and
$\bbeta/\|\bbeta\|= (0.7682,0.3841,\break 0.5121, 0,\ldots,0)^T$ is a
$p\times1$ vector, and a sample of $(X_1,\ldots,X_p)^T$ with size
$n$ is generated from a multivariate normal distribution
$N(0,\Sigma)$ whose covariance matrix $\Sigma=(\sigma_{ij})_{p\times
p}$ has entries $\sigma_{ii}=1,i=1,\ldots,p$, and
$\sigma_{ij}=\rho,i\neq j$. The replication time is again 200, and
$p=100,1000$, $n=20,50,70$ and $\rho=0,0.1,0.5,0.9$,
respectively.
We also compare the proposed method with the generalized correlation
rank method (gcorr) proposed by \citet{HalMil09} for the
logarithm transformation model (the results for the Box--Cox
transformation model are similar).

\begin{sidewaystable}
\tablewidth=\textheight
\tablewidth=\textwidth
\caption{Proportion of SIS, RRCS and IRRCS
that include the true model for the Box--Cox transformation model
$\{X_{1},X_{2},X_{3}\}$}\label{tab4}
\begin{tabular*}{\tablewidth}{@{\extracolsep{\fill}}ld{1.2}c
d{1.3}d{1.3}d{1.3}d{1.4}
d{1.3}d{1.3}d{1.3}d{1.4}
d{1.3}d{1.3}d{1.3}d{1.3}@{}}
\hline
&& \multicolumn{1}{c}{$\bolds{\varepsilon\sim}$} &
\multicolumn{4}{c}{$\bolds{N(0,1)}$} & \multicolumn{4}{c}{$\bolds{N(0,1)}$
\textbf{with 10\% outliers}} & \multicolumn{4}{c@{}}{$\bolds{t(3)}$}\\[-4pt]
&& \multicolumn{1}{c}{\hrulefill} &
\multicolumn{4}{c}{\hrulefill} & \multicolumn{4}{c}{\hrulefill}
& \multicolumn{4}{c@{}}{\hrulefill}\\
\multicolumn{1}{@{}l}{$\bolds{(p,n)}$}
&\multicolumn{1}{c}{$\bolds{\lambda}$}
&\multicolumn{1}{c}{\textbf{Method}}
& \multicolumn{1}{c}{$\bolds{\rho=0}$}
& \multicolumn{1}{c}{$\bolds{0.1}$} & \multicolumn{1}{c}{$\bolds{0.5}$}
& \multicolumn{1}{c}{$\bolds{0.9}$} & \multicolumn{1}{c}{$\bolds{0}$}
& \multicolumn{1}{c}{$\bolds{0.1}$} & \multicolumn{1}{c}{$\bolds{0.5}$}
& \multicolumn{1}{c}{$\bolds{0.9}$} & \multicolumn{1}{c}{$\bolds{0}$}
& \multicolumn{1}{c}{$\bolds{0.1}$} & \multicolumn{1}{c}{$\bolds{0.5}$} &
\multicolumn{1}{c@{}}{$\bolds{0.9}$}\\
\hline
%
%
$(100,20)$ & 0.75 &SIS & 0.415 & 0.470 & 0.190 & 0.030 &0.380&0.435&0.170 &
0.005 &
0.420 & 0.525 & 0.355 & 0.200 \\
& & RRCS & 0.440 & 0.525 & 0.400 & 0.225 &0.430&0.510&0.370 &
0.220 & 0.525
& 0.555 & 0.450 & 0.220 \\
& & IRRCS & 0.985 & 0.975 & 0.975 & 0.850 &0.940&0.910&0.875 &0.755 &
0.960 &
0.945 & 0.925 & 0.840 \\
[4pt]
& 0.5 &SIS & 0.320 & 0.390 & 0.155 & 0.005 &0.265&0.345&0.160 & 0.005 &
0.360 & 0.490
& 0.325 & 0.090 \\
& & RRCS & 0.435 & 0.525 & 0.400 & 0.225 &0.450&0.510&0.390 & 0.195
& 0.590
& 0.545 & 0.355 & 0.225 \\
& & IRRCS & 0.985 & 0.970 & 0.945 & 0.860 &0.900&0.890&0.885 &0.745 &
0.935 &
0.920 & 0.910 & 0.815
\\
[4pt]
& 0.25 &SIS & 0.150 & 0.195 & 0.090 & 0.0025&0.145&0.155&0.085 & 0.0015&
0.190 & 0.225
& 0.175 & 0.005 \\
& & RRCS & 0.435 & 0.535 & 0.395 & 0.225 &0.425&0.495&0.365 &
0.220 & 0.560
& 0.440 & 0.385 & 0.185 \\
& & IRRCS & 0.975 & 0.985 & 0.960 & 0.845 &0.905&0.885&0.870 & 0.680 &
0.910 &
0.915 & 0.895 &
0.785\\
[8pt]
$(100,50)$ & 0.75 &SIS & 0.935 & 0.915& 0.855 & 0.415 &0.875&0.905&0.795 &
0.385 &
0.890 & 0.910 & 0.850 & 0.850 \\
& & RRCS & 0.965 & 0.985& 0.955 & 0.890 &0.965&0.985&0.945 & 0.870
& 0.960
& 0.985 & 0.910 & 0.875 \\
& & IRRCS & 1 & 1 & 1 & 0.980 & 1 & 1 &0.965 & 0.925 & 1 & 1 & 0.960 & 0.910
\\
[4pt]
& 0.5 &SIS & 0.935 & 0.905& 0.810 & 0.390 &0.795&0.845&0.740 & 0.355 &
0.855 & 0.890
& 0.730 & 0.380 \\
& & RRCS & 0.965 & 0.985& 0.950 & 0.890 &0.950&0.980&0.950 & 0.880
& 0.955 &
0.940 & 0.930 & 0.840 \\
& & IRRCS & 1 & 1 & 1 & 0.980 & 1 & 1 &0.955 & 0.915 & 1 & 1 & 0.955 & 0.930
\\
[4pt]
& 0.25 &SIS & 0.815 & 0.880& 0.680 & 0.305 &0.680&0.740&0.585 & 0.260 &
0.760 & 0.860
& 0.720 & 0.370 \\
& & RRCS & 0.965 & 0.985& 0.955 & 0.900 &0.955&0.985&0.955 & 0.885
& 0.900
& 0.985 & 0.945 & 0.865 \\
& & IRRCS & 1 & 1 & 1 & 0.970 & 1 & 1 &0.975 & 0.915 & 1 & 1 & 0.985 &
0.910\\
\hline
\end{tabular*}
\end{sidewaystable}
\setcounter{table}{2}
\begin{sidewaystable}
\tablewidth=\textheight
\tablewidth=\textwidth
\caption{(Continued)}
\begin{tabular*}{\tablewidth}{@{\extracolsep{\fill}}ld{1.2}c
d{1.3}d{1.3}d{1.3}d{1.4}
d{1.3}d{1.3}d{1.3}d{1.3}
d{1.3}d{1.3}d{1.3}d{1.4}@{}}
\hline
&& \multicolumn{1}{c}{$\bolds{\varepsilon\sim}$} &
\multicolumn{4}{c}{$\bolds{N(0,1)}$} & \multicolumn{4}{c}{$\bolds{N(0,1)}$
\textbf{with 10\% outliers}} & \multicolumn{4}{c@{}}{$\bolds{t(3)}$}\\[-4pt]
&& \multicolumn{1}{c}{\hrulefill} &
\multicolumn{4}{c}{\hrulefill} & \multicolumn{4}{c}{\hrulefill}
& \multicolumn{4}{c@{}}{\hrulefill}\\
\multicolumn{1}{@{}l}{$\bolds{(p,n)}$}
&\multicolumn{1}{c}{$\bolds{\lambda}$}
&\multicolumn{1}{c}{\textbf{Method}}
& \multicolumn{1}{c}{$\bolds{\rho=0}$}
& \multicolumn{1}{c}{$\bolds{0.1}$} & \multicolumn{1}{c}{$\bolds{0.5}$}
& \multicolumn{1}{c}{$\bolds{0.9}$} & \multicolumn{1}{c}{$\bolds{0}$}
& \multicolumn{1}{c}{$\bolds{0.1}$} & \multicolumn{1}{c}{$\bolds{0.5}$}
& \multicolumn{1}{c}{$\bolds{0.9}$} & \multicolumn{1}{c}{$\bolds{0}$}
& \multicolumn{1}{c}{$\bolds{0.1}$} & \multicolumn{1}{c}{$\bolds{0.5}$} &
\multicolumn{1}{c@{}}{$\bolds{0.9}$}\\
\hline
$(1000,50)$& 0.75 &SIS & 0.615 & 0.605& 0.145 & 0 &0.515&0.490&0.130 & 0 &
0.530 &
0.570 & 0.130 & 0.005 \\
& & RRCS & 0.750 & 0.705& 0.485 & 0.230 &0.640&0.650&0.435 &0.215
& 0.710 &
0.640 & 0.435 & 0.180 \\
& & IRRCS & 1 & 1 & 1 & 0.840 &0.940&0.925&0.940 & 0.780 & 0.930 &
0.940 &
0.935 & 0.710 \\
[4pt]
& 0.5 &SIS & 0.490 & 0.510& 0.110 & 0 &0.366&0.370&0.080 & 0 & 0.455 &
0.390 & 0.150
& 0 \\
& & RRCS & 0.760 & 0.705& 0.465 & 0.245 &0.735&0.655&0.440 & 0.215
& 0.745 &
0.625 & 0.430 & 0.170 \\
& & IRRCS & 1 & 1 & 1 & 0.815 &0.950&0.920&0.930 & 0.770 & 0.975 &
0.965 &
0.940 & 0.745 \\
[4pt]
& 0.25 &SIS & 0.200 & 0.215& 0.035 & 0 &0.145&0.160&0.020 & 0 & 0.155 &
0.210 & 0.055
& 0 \\
& & RRCS & 0.755 & 0.695& 0.470 & 0.240 &0.675&0.665&0.440 & 0.215
& 0.755
& 0.615 & 0.375 & 0.215 \\
& & IRRCS & 1 & 1 & 1 & 0.780 &0.945&0.930&0.940 & 0.720 & 0.955 &
0.930 &
0.935 &
0.725\\
[8pt]
$(1000,70)$& 0.75 &SIS & 0.860 & 0.860& 0.375 & 0.005 &0.670&0.690&0.270
&0.015 & 0.840
& 0.865 & 0.370 & 0.105 \\
& & RRCS & 0.880 & 0.890& 0.725 & 0.515 &0.880&0.880&0.695 &0.510
& 0.915 &
0.885 & 0.700 & 0.395 \\
& & IRRCS & 1 & 1 & 1 & 0.970 &0.960&0.945&0.935
&0.910 & 0.970 & 0.985 & 0.930 & 0.915\\
[4pt]
& 0.5 &SIS & 0.775 & 0.765& 0.275 & 0.0015&0.555&0.585&0.230 & 0 & 0.760 &
0.750 &
0.280 & 0.0015 \\
& & RRCS & 0.885 & 0.900& 0.715 & 0.470 &0.865&0.875&0.670 &0.515 &
0.915 &
0.875 & 0.610 & 0.440 \\
& & IRRCS & 1 & 1 & 1 & 0.950 &0.955&0.945&0.935 &0.900 & 0.955 & 0.950
& 0.915
& 0.875 \\
[4pt]
& 0.25 &SIS & 0.435 & 0.445& 0.010 & 0 &0.365&0.290&0.075 & 0 & 0.440 &
0.440 & 0.010
& 0 \\
& & RRCS & 0.875 & 0.880& 0.725 & 0.490 &0.830&0.795&0.710 & 0.500
& 0.835
& 0.830 & 0.655 & 0.410 \\
& & IRRCS & 1 & 1 & 1 & 0.920 &0.960&0.940&0.935 & 0.900 & 0.955 &
0.935 &
0.925 &
0.885\\
\hline
\end{tabular*}
\end{sidewaystable}

\begin{sidewaystable}
\tablewidth=\textheight
\tablewidth=\textwidth
\caption{Proportion of SIS, gcorr, RRCS and
IRRCS that include the true model for the logarithm transformation
model}\label{tab5}
\begin{tabular*}{\tablewidth}{@{\extracolsep{\fill}}lcd{1.3}d{1.3}d{1.3}d{1.3}
d{1.3}d{1.3}d{1.3}d{1.3}d{1.3}d{1.3}d{1.3}d{1.3}@{}}
\hline
& \multicolumn{1}{c}{$\bolds{\varepsilon\sim}$} &
\multicolumn{4}{c}{$\bolds{N(0,1)}$} & \multicolumn{4}{c}{$\bolds{N(0,1)}$
\textbf{with 10\% outliers}} & \multicolumn{4}{c@{}}{$\bolds{t(3)}$}\\[-4pt]
& \multicolumn{1}{c}{\hrulefill} &
\multicolumn{4}{c}{\hrulefill} & \multicolumn{4}{c}{\hrulefill}
& \multicolumn{4}{c@{}}{\hrulefill}\\
\multicolumn{1}{@{}l}{$\bolds{(p,n)}$}
&\multicolumn{1}{c}{\textbf{Method}}
& \multicolumn{1}{c}{$\bolds{\rho=0}$}
& \multicolumn{1}{c}{$\bolds{0.1}$} & \multicolumn{1}{c}{$\bolds{0.5}$}
& \multicolumn{1}{c}{$\bolds{0.9}$} & \multicolumn{1}{c}{$\bolds{0}$}
& \multicolumn{1}{c}{$\bolds{0.1}$} & \multicolumn{1}{c}{$\bolds{0.5}$}
& \multicolumn{1}{c}{$\bolds{0.9}$} & \multicolumn{1}{c}{$\bolds{0}$}
& \multicolumn{1}{c}{$\bolds{0.1}$} & \multicolumn{1}{c}{$\bolds{0.5}$} &
\multicolumn{1}{c@{}}{$\bolds{0.9}$}\\
\hline
$(100,20)$ & SIS & 0.100 & 0.060 & 0.070 & 0.030 & 0.055 & 0.065 &0.020 &
0.020 &
0.040 & 0.060 & 0.030 & 0.015 \\
& gcorr & 0.280 & 0.230 & 0.105 & 0.010 & 0.205 & 0.215 &0.180 & 0.010
& 0.185 &
0.230 & 0.170 & 0.015 \\
&RRCS & 0.580 & 0.460 & 0.385 & 0.290 & 0.570 & 0.410 &0.375 & 0.215 &
0.575 &
0.425 & 0.355 & 0.170 \\
& IRRCS & 1 & 0.975 & 0.975 & 0.715 & 0.875 & 0.870 &0.875 & 0.560 &
0.905 &
0.875 & 0.840 & 0.580 \\
[4pt]
$(100,50)$ & SIS & 0.550 & 0.650 &0.450 &0.225 & 0.470 & 0.585 &0.395 &
0.250 &
0.470 & 0.585 & 0.455 & 0.230 \\
& gcorr & 0.940 & 0.925 & 0.890 &0.430 & 0.855 & 0.880 &0.825 & 0.385 &
0.870 &
0.885 & 0.860 & 0.410 \\
&RRCS & 0.960 & 0.985 &0.975 & 0.880 & 0.960 & 0.975 &0.965 & 0.930 &
0.985 &
0.975 & 0.945 & 0.865 \\
& IRRCS & 1 & 1 & 1 & 0.980 & 1 & 1 & 1 &
0.955 & 0.990 & 1 & 1 & 0.975\\
[8pt]
$(1000,50)$& SIS & 0.035 & 0.020 & 0.005 & 0 & 0.015 & 0.005 &0.020 &
0.010 &
0.020 & 0.010 & 0.005 & 0 \\
& gcorr & 0.420 & 0.415 & 0.285 & 0.015 & 0.385 & 0.405 &0.025 & 0.005
& 0.340 &
0.410 & 0.265 & 0.010 \\
&RRCS & 0.610 & 0.670 &0.490 & 0.225 & 0.630 & 0.590 &0.400 & 0.200 &
0.605 &
0.650 & 0.495 & 0.155 \\
& IRRCS & 1 & 1 & 1 & 0.855 & 0.925 & 0.900 &0.915 & 0.685 & 1 & 1 &
0.990 &
0.660 \\
[4pt]
$(1000,70)$& SIS & 0.125 &0.080 & 0.005 & 0 &0.075 & 0.040 &0.005 & 0 &
0.080 &
0.055 & 0.010 & 0.005 \\
& gcorr & 0.695 &0.640 & 0.615 &0.230 &0.625 & 0.630 &0.440 & 0.185 &
0.590 &
0.625 & 0.480 & 0.205 \\
&RRCS & 0.915 &0.845 & 0.785 &0.475 &0.870 & 0.880 &0.665 & 0.485 &
0.860 & 0.840
& 0.650 & 0.450 \\
& IRRCS & 1 & 1 & 1 & 0.940 & 1 & 1 &0.960 &
0.930 & 1 & 1 & 1 & 0.925\\
\hline
\end{tabular*}
\end{sidewaystable}

From Tables~\ref{tab4} and~\ref{tab5}, we can see clearly that
without exception RRCS outperforms SIS and gcorr significantly and
IRRCS can greatly improve the performance of RRCS.
%
%
%
%
\end{Exam}

\begin{Exam}[(Logistic regression)]\label{Example4}
In this example, the data $(\bfX_1^T, Y_1),\ldots,\break(\bfX_n^T,Y_n)$ are
independent copies of a pair $(\bfX^T, Y)$, where the conditional
distribution of the response $Y$ given $X$ is a binomial distribution
with
%
\begin{equation}
\label{logistic} \log\biggl(\frac{p(\bfX)}{1-p(\bfX)} \biggr)=\bfX
^{T}\bbeta.
\end{equation}
The predictors are generated in the same setting as that of
\citet{FanSon10}, that is,
\[
X_j=\frac{\varepsilon_j+a_j\varepsilon}{\sqrt{1+a_j^2}},
\]
where $\varepsilon$ and $\{\varepsilon_j\}_{j=1}^{[p/3]}$ are i.i.d.
standard normal, $\{\varepsilon_j\}_{j=[p/3]+1}^{[2p/3]}$ are i.i.d.
and follow a double exponential distribution with location parameter
zero and scale parameter one, and
$\{\varepsilon_j\}_{j=[2p/3]+1}^{[p]}$ are i.i.d.\vspace*{1pt} and
follow a mixture normal distribution with two components $N(-1, 1),
N(1, 0.5)$ and equal mixture proportion. The predictors are
standardized to be mean zero and variance one. The
constants\vspace*{1pt} $\{a_j\}_{j=1}^q$ are the same and chosen such
that the correlation $\rho= \operatorname{corr}(X_i,X_j) = 0, 0.2, 0.4,
0.6$ and 0.8, among the first $q$ predictors, and $a_j = 0$ for $j >q$.
Parameter $q$ is also related to the overall correlation in the
covariance matrix.

We vary the size of the nonsparse set of coefficients as $s = 3, 6, 12,
15$ and~24, and present the numerical results with $q=15$ and $q=50$.
Every method is evaluated by summarizing the median minimum model size
(MMMS) of the selected model and its associated RSD, which is the
associated interquartile range (IQR) divided by 1.34. The results,
based on 200 replications in each scenario, are recorded in Tables
\ref{tab6}--\ref{tab8}. The results of SIS-based MLR, SIS-based MMLE,
LASSO and SCAD in Tables~\ref{tab6}--\ref{tab8} are cited from
\citet{FanSon10}.

\begin{sidewaystable}
\tablewidth=\textheight
\tablewidth=\textwidth
\caption{The MMMS and associated RSD (in
parenthesis) of the simulated examples for logistic regressions when
$p = 40\mbox{,}000$}\label{tab6}
{\fontsize{8.9pt}{10.5pt}\selectfont{
\begin{tabular*}{\tablewidth}{@{\extracolsep{\fill}}lck{3.5}k{5.5}
k{5.6}ck{5.5}k{5.6}k{5.8}@{\hspace*{-1pt}}}
\hline
$\bolds{\rho}$& \multicolumn{1}{c}{$\bolds{n}$}
& \multicolumn{1}{c}{\textbf{SIS-MLR}} &
\multicolumn{1}{c}{\textbf{SIS-MMLE}}
& \multicolumn{1}{c}{\textbf{RRCS}} & \multicolumn{1}{c}{$\bolds{n}$}
& \multicolumn{1}{c}{\textbf{SIS-MLR}} &
\multicolumn{1}{c}{\textbf{SIS-MMLE}} &
\multicolumn{1}{c@{}}{\textbf{RRCS}}\\
\hline
&\multicolumn{8}{c@{}}{Setting 1, $q=15$} \\
&\multicolumn{4}{c}{$s=3, \bbeta=(1,1.3,1)^T$}
&\multicolumn{4}{c@{}}{$s=6, \bbeta=(1,1.3,1,\ldots)^T$} \\
[3pt]
0 &300 &3\sset{1} & 3\sset{1} & 3\sset{0.74} & 300 &47\sset{164} &50\sset{170} & 56\sset{188.05} \\
0.2 &200 &3\sset{0} &3\sset{0} & 3\sset{0} & 300 & 6\sset{0} & 6\sset{0} & 6\sset{0.74} \\
0.4 &200 &3\sset{0} &3\sset{0} & 3\sset{0} & 300 & 7\sset{1} & 7\sset{1} & 7\sset{1.49} \\
0.6 &200 &3\sset{1} &3\sset{1} & 3\sset{0.74} & 300 & 8\sset{1} & 8\sset{2} & 8\sset{2.23} \\
0.8 &200 &4\sset{1} &4\sset{1} & 4\sset{2} & 300 & 9\sset{3} & 9\sset{3} & 9\sset{2.23} \\
[3pt]
&\multicolumn{4}{c}{$s=12, \bbeta=(1,1.3,\ldots)^T$} &\multicolumn
{4}{c@{}}{$s=15, \bbeta=(1,1.3,\ldots)^T$} \\
[3pt]
0 &500 &297\sset{589} &302.5\sset{597}&298\sset{488}& 600 &350\sset{607}&359.5\sset{612}&
359.5\sset{657.08} \\
0.2 &300 &13\sset{1} &13\sset{1} &13\sset{1.49}& 300 & 15\sset{0} & 15\sset{0} & 15\sset{0} \\
0.4 &300 &14\sset{1} &14\sset{1} &14\sset{0.74}& 300 & 15\sset{0} & 15\sset{0} & 15\sset{0} \\
0.6 &300 &14\sset{1} &14\sset{1} &14\sset{1.49}& 300 & 15\sset{0} & 15\sset{0} & 15\sset{0} \\
0.8 &300 &14\sset{1} &14\sset{1} &14\sset{0.74}& 300 & 15\sset{0} & 15\sset{0}
& 15\sset{0} \\
[6pt]
&\multicolumn{8}{c@{}}{Setting 2, $q=50$} \\
&\multicolumn{4}{c}{$s=3, \bbeta=(1,1.3,1)^T$}
&\multicolumn{4}{c@{}}{$s=6, \bbeta=(1,1.3,1,\ldots)^T$} \\
[3pt]
0 &300 &3\sset{1} & 3\sset{1}& 3\sset{0.74}& 500 & 6\sset{1} & 6\sset{1} & 6\sset{2} \\
0.2 &300 &3\sset{0} &3\sset{0} & 3\sset{0} & 500 & 6\sset{0} & 6\sset{0} & 6\sset{0} \\
0.4 &300 &3\sset{0} &3\sset{0} & 3\sset{0} & 500 & 6\sset{1} & 6\sset{1} & 7\sset{1.49} \\
0.6 &300 &3\sset{1} &3\sset{1} & 3\sset{1} & 500 & 8.5\sset{4} & 9\sset{5} & 8\sset{3.73} \\
0.8 &300 &5\sset{4} &5\sset{4} & 5\sset{3.73}& 500 & 13.5\sset{8}& 14\sset{8} & 15\sset{7.46}\\
[3pt]
&\multicolumn{4}{c}{$s=12, \bbeta=(1,1.3,\ldots)^T$} &
\multicolumn{4}{c@{}}{$s=15, \bbeta=(1,1.3,\ldots)^T$} \\
[3pt]
0 &600 &77\sset{114} & 78.5\sset{118}& 95\sset{115} & 800 &46\sset{82} & 47\sset{83}& 46\sset{83.88}\\
0.2 &500 &18\sset{7} &18\sset{7} & 19\sset{6} & 500 & 26\sset{6} & 26\sset{6} & 27\sset{8.20} \\
0.4 &500 &25\sset{8} &25\sset{10} & 26\sset{9.70} & 500 & 34\sset{7} & 33\sset{8} & 33\sset{8.39} \\
0.6 &500 &32\sset{9} &31\sset{8} & 32\sset{9} & 500 & 39\sset{7} & 38\sset{7} & 38\sset{6.71} \\
0.8 &500 &36\sset{8} &35\sset{9} & 39\sset{7.46} & 500 & 40\sset{6} &
42\sset{7} & 42\sset{6.15}\\
\hline
\end{tabular*}}}
\end{sidewaystable}

From Tables~\ref{tab6}--\ref{tab8}, we can see that the RRCS
procedure does a very reasonable job similar to the SIS proposed by
\citet{FanSon10} in screening insignificant predictors, and
similarly sometimes outperforms LASSO and SCAD for NP-dimensional
generalized linear models.
\end{Exam}

\begin{Exam}[(Logistic regression)]\label{Example5}
This example is based on a real data set
from Example 11.3 of \citet{AlbWinZap99}. This data set
consists of 208 employees with complete information on 8 recorded
variables. These variables include employee's annual salary in
thousands of dollars (Salary); educational level (EduLev), a
categorical variable with categories 1 (finished school), 2
(finished some college courses), 3 (obtained a bachelor's degree), 4
(took some graduate courses), 5 (obtained a graduate degree); job
%
\begin{table}
\caption{The MMMS and associated RSD (in
parenthesis) of the simulated examples for logistic regressions when
$p = 5000$ and $q=15$}\label{tab7}
\begin{tabular*}{\tablewidth}{@{\extracolsep{\fill}}lck{5.5}k{3.5}k{4.6}k{2.6}
k{5.7}@{\hspace*{-2pt}}}
\hline
$\bolds{\rho}$& \multicolumn{1}{c}{$\bolds{n}$}
& \multicolumn{1}{c}{\textbf{SIS-MLR}} & \multicolumn{1}{c}{\textbf{SIS-MMLE}}
& \multicolumn{1}{c}{\textbf{LASSO}} & \multicolumn{1}{c}{\textbf{SCAD}}
& \multicolumn{1}{c@{}}{\textbf{RRCS}}\\
\hline
&\multicolumn{6}{c@{}}{$s=3, \bolds{\beta}=(1,1.3,1)^T$} \\
[3pt]
0 &300 &3\sset{0} &3\sset{0} &3\sset{1} & 3\sset{1} &3\sset{0} \\
0.2 &300 &3\sset{0} &3\sset{0} &3\sset{0} & 3\sset{0} &3\sset{0} \\
0.4 &300 &3\sset{0} &3\sset{0} &3\sset{0} & 3\sset{0} &3\sset{0} \\
0.6 &300 &3\sset{0} &3\sset{0} &3\sset{0} & 3\sset{1} &3\sset{0} \\
0.8 &300 &3\sset{1} &3\sset{1} &4\sset{1} & 4\sset{1} &3\sset{1.49} \\
[3pt]
&\multicolumn{6}{c@{}}{$s=6, \bolds{\beta}=(1,1.3,1,1.3,1,1.3)^T$} \\
[3pt]
0 &300 &12.5\sset{15} &13\sset{6} &7\sset{1} & 6\sset{1} &12\sset{24.62} \\
0.2 &300 &6\sset{0} &6\sset{0} &6\sset{0} & 6\sset{0} &6\sset{0.18} \\
0.4 &300 &6\sset{1} &6\sset{1} &6\sset{1} & 6\sset{0} &7\sset{1.49} \\
0.6 &300 &7\sset{2} &7\sset{2} &7\sset{1} & 6\sset{1} &8\sset{1.49} \\
0.8 &300 &9\sset{2} &9\sset{3} &27.5\sset{3725}& 6\sset{0} &9\sset{2.23} \\
[3pt]
&\multicolumn{6}{c@{}}{$s=12, \bolds{\beta}=(1,1.3,\ldots)^T$} \\
[3pt]
0 &300 &297.5\sset{359} &300\sset{361} &72.5\sset{3704}& 12\sset{0} &345\sset{522} \\
0.2 &300 &13\sset{1} &13\sset{1} &12\sset{1} & 12\sset{0} &13\sset{1.49} \\
0.4 &300 &14\sset{1} &14\sset{1} &14\sset{1861} & 13\sset{1865} &14\sset{0.74} \\
0.6 &300 &14\sset{1} &14\sset{1} &2552\sset{85} & 12\sset{3721} &14\sset{1} \\
0.8 &300 &14\sset{1} &14\sset{1} &2556\sset{10} & 12\sset{3722} &14\sset{0.74} \\
[3pt]
&\multicolumn{6}{c@{}}{$s=15, \bolds{\beta}=(3,4,\ldots)^T$} \\
[3pt]
0 &300 &479\sset{622} &482\sset{615} &69.5\sset{68} & 15\sset{0} &629.5\sset{821} \\
0.2 &300 &15\sset{0} &15\sset{0} &16\sset{13} & 15\sset{0} &15\sset{0} \\
0.4 &300 &15\sset{0} &15\sset{0} &38\sset{3719} & 15\sset{3720} &15\sset{0} \\
0.6 &300 &15\sset{0} &15\sset{0} &2555\sset{87} & 15\sset{1472} &15\sset{0} \\
0.8 &300 &15\sset{0} &15\sset{0} &2552\sset{8} & 15\sset{1322} &15\sset{0}\\
\hline
\end{tabular*}
\end{table}
grade (JobGrade), a categorical variable indicating the current job
level, the possible levels being 1--6 (6~the highest); year that an
employee was hired (YrHired); year that an employee was born
(YrBorn); a categorical variable with values ``Female'' and ``Male''
(Gender), 1 for female employee and 0 for male employee; number of
years of work experience at another bank prior to working at the
%
\begin{table}
\caption{The MMMS and associated RSD (in
parenthesis) of the simulated examples for logistic regressions when
$p = 2000$ and $q=50$}
\label{tab8}
\begin{tabular*}{\tablewidth}{@{\extracolsep{\fill}}lck{5.5}
k{4.5}k{4.4}k{4.6}k{4.8}@{\hspace*{-2pt}}}
\hline
$\bolds{\rho}$& \multicolumn{1}{c}{$\bolds{n}$}
& \multicolumn{1}{c}{\textbf{SIS-MLR}} & \multicolumn{1}{c}{\textbf{SIS-MMLE}}
& \multicolumn{1}{c}{\textbf{LASSO}} & \multicolumn{1}{c}{\textbf{SCAD}}
& \multicolumn{1}{c@{}}{\textbf{RRCS}}\\
\hline
&\multicolumn{6}{c}{$s=3, \bolds{\beta}=(3,4,3)^T$} \\
[3pt]
0 &200 &3\sset{0} &3\sset{0} &3\sset{0} & 3\sset{0} &3\sset{0} \\
0.2 &200 &3\sset{0} &3\sset{0} &3\sset{0} & 3\sset{0} &3\sset{0} \\
0.4 &200 &3\sset{0} &3\sset{0} &3\sset{0} & 3\sset{1} &3\sset{0} \\
0.6 &200 &3\sset{1} &3\sset{1} &3\sset{1} & 3\sset{1} &3\sset{0.74} \\
0.8 &200 &5\sset{5} &5.5\sset{5} &6\sset{4} & 6\sset{4} &4\sset{2.4} \\
[3pt]
&\multicolumn{6}{c}{$s=6, \bolds{\beta}=(3,-3,3,-3,3,-3)^T$} \\
[3pt]
0 &200 &8\sset{6} &9\sset{7} &7\sset{1} & 7\sset{1} &8\sset{5.97} \\
0.2 &200 &18\sset{38} &20\sset{39} &9\sset{4} & 9\sset{2} &14\sset{28.54} \\
0.4 &200 &51\sset{77} &64.5\sset{76} &20\sset{10} & 16.5\sset{6} &72\sset{76.60} \\
0.6 &300 &77.5\sset{139} &77.5\sset{132} &20\sset{13} & 19\sset{9} &84.5\sset{122.94} \\
0.8 &400 &306.5\sset{347} &313\sset{336} &86\sset{40} & 70.5\sset{35} &249.5\sset{324.62} \\
[3pt]
&\multicolumn{6}{c}{$s=12, \bolds{\beta}=(3,4,\ldots)^T$} \\
[3pt]
0 &600 &13\sset{6} &13\sset{7} &12\sset{0} & 12\sset{0} &13\sset{3.90} \\
0.2 &600 &19\sset{6} &19\sset{6} &13\sset{1} & 13\sset{2} &16.5\sset{4} \\
0.4 &600 &32\sset{10} &30\sset{10} &18\sset{3} & 17\sset{4} &23\sset{7} \\
0.6 &600 &38\sset{9} &38\sset{10} &22\sset{3} & 22\sset{4} &29\sset{8.95} \\
0.8 &600 &38\sset{7} &39\sset{8} &1071\sset{6} & 1042\sset{34} &35\sset{8} \\
[3pt]
&\multicolumn{6}{c}{$s=24, \bolds{\beta}=(3,4,\ldots)^T$} \\
[3pt]
0 &600 &180\sset{240} &182\sset{238} &35\sset{9} & 31\sset{10} &190.5\sset{240.48} \\
0.2 &600 &45\sset{4} &45\sset{4} &35\sset{27} & 32\sset{24} &40\sset{5} \\
0.4 &600 &46\sset{3} &47\sset{2} &1099\sset{17} & 1093\sset{1456} &45\sset{4.40} \\
0.6 &600 &48\sset{2} &48\sset{2} &1078\sset{5} & 1065\sset{23} &47\sset{3} \\
0.8 &600 &48\sset{1} &48\sset{1} &1072\sset{4} & 1067\sset{13} &47\sset{2.98}\\
\hline
\end{tabular*}
\end{table}
Fifth National Bank (YrsPrior); a~dummy variable with value~1 if the
employee's job is computer related and value 0 otherwise (PCJob).\vadjust{\goodbreak}
Such a data set had been analyzed by \citet{FanPen04} throughout
the following linear model:
%
\begin{eqnarray}
\label{bank-M}
\mathrm{Salary}&=&\beta_0+\beta_{1}\mathrm{Female}+
\beta_{2}\mathrm{PCJob}
+\sum_{i=1}^{4}
\beta_{2+i}\mathrm{Edu}_{i}
+\sum_{i=1}^{5}
\beta_{6+i}\mathrm{JobGrd}_{i}\nonumber\\[-8pt]\\[-8pt]
&&{}+\beta_{12}\mathrm{YrsExp}+\beta_{13}
\mathrm{Age}+\varepsilon,\nonumber
\end{eqnarray}
where the variable YrsExp is total years of working experience,
computed from the variables YrHired and YrsPrior. \citet{FanPen04}
deleted the samples with age over 60 or working experience over 30 and
used only\ 199 samples to fit model~(\ref{bank-M}). The SCAD-penalized
least squares coefficient estimator of (\ref{bank-M}) is
\begin{eqnarray*}
\bbeta_0 &=& (\beta_0,\beta_1,\ldots,
\beta_{13})^T
\\
&=& (55.835,-0.624,4.151,0, -1.073,-0.914,0,-24.643,
\\
&&\hspace*{43pt} -22.818, -18.803, -13.859,-7.770,0.193,0)^T.
\end{eqnarray*}
For this data set, we consider a larger artificial model as a full
model with additional predictors: 
%
\[
Y_j={\beta}_0+\sum
_{i=1}^{13} \beta_{i}X_{ij}+\sum
_{i=14}^{[2p/5]}\beta_{i}X_{ij}+
\sum_{[2p/5]+1}^{p}\beta_{i}X_{ij}+
\sigma\varepsilon_j,\qquad j=1,\ldots,n,
\]
where we set $({\beta}_0,{\beta}_1,\ldots,{\beta}_{13})^T=\bbeta_0$
that is identical to that of (\ref{bank-M}) above by
\citet{FanPen04}, and set $\beta_i=0$, for $i$ with $13<i \le p$.
Hence, $X_{3j}, X_{6j}$, $X_{13j}$ and~$X_{ij}$, $13<i\le p$, are
insignificant covariates, whose corresponding coefficients are zero.
The data are generated as follows. $(X_{1j},\ldots, X_{13j},
j=1,\ldots,n)$ are corresponding to the covariates in (\ref{bank-M})
and resampled from those 199 real data without replacement. For each
$i$, $X_{ij}, 14 \le i \le [2p/5]$, are\vadjust{\goodbreak} generated independently from
the Bernoulli distribution with success probability $p^\ast_i$ where
$p^\ast_i$ is independently random sampled from the uniform
distribution under the interval $[0.2,0.8]$, and $X_{ij}, [2p/5]+1 \le
i \le p$, are generated independently from the standard normal
distribution. Further, the noises $\varepsilon_j, 1 \le j \le n$, are,
respectively, generated from the normal distribution with zero mean and
the standard error $\sigma=1,2,3$.

To compare the performance of different methods, we set the sample
size $n$ to be 180, and, respectively, consider the different dimensions
$p=200, 400,
600$ and 1000. Consider the different sizes of $d_n=15,30,60,120$
and 179 predictors for the sure screening by the three different
methods: RRCS, SIS and the generalized correlation rank method
(gcorr) proposed by \citet{HalMil09}. Then we compute the
proportion of the models that include the true one, which are selected
by RRCS, SIS and gcorr, respectively. The
experiment is repeated 200 times and the results are reported in
Table~\ref{bank} for various combinations of $p$ and $d_n$.

\begin{sidewaystable}
\tablewidth=\textheight
\tablewidth=\textwidth
\caption{For Example \protect\ref{Example5}: the proportion of
RRCS, SIS and gcorr that include the true model}\label{bank}
\begin{tabular*}{\tablewidth}{@{\extracolsep{\fill}}lcd{1.3}
d{1.3}d{1.3}d{1.3}d{1.3}d{1.3}d{1.3}d{1.3}d{1.3}d{1.3}d{1.3}d{1.3}@{}}
\hline
&&\multicolumn{4}{c}{$\bolds{\sigma=1}$}&\multicolumn{4}{c}{$\bolds{\sigma
=2}$}&\multicolumn{4}{c@{}}{$\bolds{\sigma=3}$}\\[-4pt]
&&\multicolumn{4}{c}{\hrulefill} &\multicolumn{4}{c}{\hrulefill}
&\multicolumn{4}{c@{}}{\hrulefill}\\
$\bolds{d_n}$ & \textbf{Method} &
\multicolumn{1}{c}{$\bolds{p=200}$} &
\multicolumn{1}{c}{\textbf{400}} & \multicolumn{1}{c}{\textbf{600}}
& \multicolumn{1}{c}{\textbf{1000}} & \multicolumn{1}{c}{\textbf{200}}
& \multicolumn{1}{c}{\textbf{400}} &\multicolumn{1}{c}{\textbf{600}}
& \multicolumn{1}{c}{\textbf{1000}} & \multicolumn{1}{c}{\textbf{200}} &
\multicolumn{1}{c}{\textbf{400}} & \multicolumn{1}{c}{\textbf{600}}
& \multicolumn{1}{c@{}}{\textbf{1000}}\\
\hline
\hphantom{0}15& RRCS & 0.280 & 0.080 & 0& 0 & 0.085 & 0 & 0 & 0 & 0.005 & 0 & 0 & 0
\\
& SIS & 0 & 0 & 0 & 0 & 0 & 0 & 0 & 0 & 0 & 0 & 0 & 0 \\
& gcorr & 0 & 0 & 0 & 0 & 0 & 0 & 0 & 0 & 0 & 0 & 0 & 0 \\
[3pt]
\hphantom{0}30& RRCS & 0.955 & 0.765 &0.425 &0.165 & 0.685 & 0.255 &0.085&0.020& 0.210
&0.030&0.005& 0 \\
& SIS & 0 & 0 & 0 & 0 & 0 & 0 & 0 & 0 & 0 & 0 & 0 & 0 \\
& gcorr & 0 & 0 & 0 & 0 & 0 & 0 & 0 & 0 & 0 & 0 & 0 & 0 \\
[3pt]
\hphantom{0}60& RRCS & 1 & 0.990 &0.915 &0.735 & 0.965 & 0.765 &0.490 &0.275& 0.620 &0.310
&0.070& 0.025\\
& SIS & 0 & 0 & 0 & 0 & 0 & 0 & 0 & 0 & 0.005 & 0 & 0 & 0 \\
& gcorr & 0 & 0 & 0 & 0 & 0 & 0 & 0 & 0 & 0 & 0 & 0 & 0 \\
[3pt]
120& RRCS & 1 & 1 & 0.995 &0.990 & 0.985 & 0.995 &0.885&0.665& 0.920
&0.670&0.410& 0.215 \\
& SIS & 0.045 & 0 & 0 & 0 & 0.070 & 0 & 0 & 0 & 0.125 &0.005& 0 & 0 \\
& gcorr & 0 & 0 & 0 & 0 & 0 & 0 & 0 & 0 & 0.050 & 0 & 0 &
0\\
[3pt]
179& RRCS & 1 & 1 & 1 & 0.995 & 1 & 1 &0.965&0.860& 0.970 &0.865&0.640& 0.410
\\
& SIS & 0.670 & 0 & 0 & 0 & 0.660 & 0.010 & 0 & 0 & 0.715 &0.015& 0 & 0
\\
& gcorr & 1 & 0 & 0 & 0 & 1 & 0 & 0 & 0 & 1 & 0 & 0 & 0 \\
\hline
\end{tabular*}
\end{sidewaystable}

From Table~\ref{bank}, we can see that the RRCS procedure works well
in screening out insignificant predictors when there are the
categorical covariates. In contrast, the SIS and gcorr methods
almost cannot choose the true model. In most of the repeated
experiments, we find that there are always one or two significant
predictors not being selected by the SIS and gcorr methods even when
$d_n=n-1=179$ predictors are selected.

For SIS, such a result is consistent with the numerical study of
Example~\ref{Example2} in \citet{FanFenSon11}. With complex correlation
structure among predictors and the response, SIS cannot work well.
As for the generalized correlation screening method, its computation
is complicated, especially because it has to use different methods to,
respectively, calculate the generalized coefficients between the
response and both categorial and continuous predictors. The
variation of those coefficient estimations would be different, and
make that the final sure screening results are not as stable as RRCS
and SIS are.\vspace*{-3pt}
\end{Exam}

%

\subsection{Application to cardiomyopathy microarray data}\label{sec5.2}
Please see the supplementary material for the paper [\citet{Lietal}].\vspace*{-3pt}

\section{Concluding remarks}\label{sec6}\label{sec7}

This paper studies the sure screening properties of robust rank
correlation screening (RRCS) for ultra-high dimensional linear
regression models and transformation regression models. The method
is based on the Kendall $\tau$ rank correlation, which is a robust
correlation measurement between two random variables and is
invariant to strictly monotonic transformation. Our results discover
the relationship between the Pearson correlation and the
Kendall $\tau$ rank correlation under certain conditions. It
suggests that the Kendall $\tau$ rank correlation can be used to
replace the Pearson correlation such that the sure screening is
applicable not only to linear regression models but also to more
general nonlinear regression models.\looseness=-1

In both the theoretical analysis and the numerical study, RRCS
has been shown to be capable\vadjust{\goodbreak} of reducing the exponentially growing
dimensionality of the model to a value smaller than the sample size.
It is also robust against the error distribution. An iterative RRCS
(IRRCS) has been also proposed to enhance the performance of RRCS
for more complicated ultra-high dimensional data.

Some issues deserve further study. From \citet{FanSon10},
it is easy to know that the sure screening properties of MMLE for
generalized linear models really depend on
$\operatorname{Cov}(X_k,Y),i=1,2,\ldots,n$. Hence, it is an interesting
problem to determine whether the relationship between the Pearson
correlation and the Kendall $\tau$ rank correlation can be
identified for generalized linear models. If this can be done, the
sure screening properties of RRCS for generalized linear models can
also be studied theoretically. Note that the conditions required
are much weaker than SIS needs. Thus, it would be of interest to
determine whether robust LASSO, SCAD or other penalized methods can
be defined when the idea described herein is applied.

\begin{appendix}\label{app}
\section*{Appendix: Proofs of theorems}

Please see the supplementary material for the paper [\citet{Lietal}].
\end{appendix}

\section*{Acknowledgments}

Most work of this paper was finished independently by the second
author or under his guidance and suggestion. The authors would like
to thank Professor Jianqing Fan for his valuable suggestions and
constructive discussion with the second author that improve the
presentation and the results of the paper. The authors would like to
thank the Editor, an Associate Editor and the referees for their
helpful comments that led to an improvement of an earlier
manuscript.

\begin{supplement}
\stitle{Supplement to ``Robust rank correlation based screening''\\}
\slink[doi]{10.1214/12-AOS1024SUPP} 
\sdatatype{.pdf}
\sfilename{aos1024\_supp.pdf}
\sdescription{Application to Cardiomyopathy
microarray Data and the proofs of Theorems~\ref{theo1}--\ref{theo3} and Proposition~\ref{prop1}
require some technical and lengthy arguments that we develop in this
supplement.}
\end{supplement}


\printaddresses

\end{document}